\documentclass[useAMS,usenatbib,twocolumn]{mn2e}
\pdfoutput=1
\usepackage{amssymb}
\usepackage{amsmath}
\usepackage{natbib}
\usepackage[pdftex]{graphicx}
\usepackage{subfigure}
\usepackage{booktabs}
\usepackage{tabularx}
\begin{document}

\title[Instability in eccentric discs]
{Hydrodynamic instability in eccentric astrophysical discs}
      \author[A. J. Barker \& G. I. Ogilvie]{A. J. Barker\thanks{Email address: ajb268@cam.ac.uk} and G. I. Ogilvie\\
Department of Applied Mathematics and Theoretical Physics, University of Cambridge, Centre for Mathematical Sciences, \\ Wilberforce Road, Cambridge CB3 0WA, UK}
	
\pagerange{\pageref{firstpage}--\pageref{lastpage}} \pubyear{2014}

\maketitle

\label{firstpage}

\begin{abstract}
Eccentric Keplerian discs are believed to be unstable to three-dimensional hydrodynamical instabilities driven by the time-dependence of fluid properties around an orbit. These instabilities could lead to small-scale turbulence, and ultimately modify the global disc properties. We use a local model of an eccentric disc, derived in a companion paper, to compute the nonlinear vertical (``breathing mode") oscillations of the disc. We then analyse their linear stability to locally axisymmetric disturbances for any disc eccentricity and eccentricity gradient using a numerical Floquet method. In the limit of small departures from a circular reference orbit, the instability of an isothermal disc is explained analytically. We also study analytically the small-scale instability of an eccentric neutrally stratified polytropic disc with any polytropic index using a WKB approximation. We find that eccentric discs are generically unstable to the parametric excitation of small-scale inertial waves. The nonlinear evolution of these instabilities should be studied in numerical simulations, where we expect them to lead to a decay of the disc eccentricity and eccentricity gradient as well as to induce additional transport and mixing. Our results highlight that it is essential to consider the three-dimensional structure of eccentric discs, and their resulting vertical oscillatory flows, in order to correctly capture their evolution.
\end{abstract}


\section{Introduction}
Astrophysical discs with eccentric orbits have been proposed to explain a number of astrophysical observations. They are thought to explain the superhump phenomenon in SU UMa stars \citep{Whitehurst1988,Lubow1991a,Smith2007}, the spectral variability of rapidly rotating Be stars \citep{Okazaki1991,PapSav1992,Ogilvie2008} and, in the case of a collisionless disc of stars, the visible structure of the nucleus of the galaxy M31 \citep{Tremaine1995,PeirisTremaine2003}. In addition, the orbital evolution of a newly born planet due to its tidal interaction with the protoplanetary disc is intricately coupled with the evolution of eccentric modes in the disc. The excitation and damping of these modes may have played a role in the early evolution of planetary eccentricities \citep{PapNM2001,Pap2002,GoldreichSari2003,KleyDirksen2006,DAngelo2006,Bitsch2013}.

Eccentric modes in Keplerian discs are slowly precessing modes with azimuthal wavenumber $m=1$ (e.g.~\citealt{Tremaine2001,Pap2002}) 
that vary on a length scale comparable with the radius of the disc. Because of their global extent they are usually thought to be extremely long-lived. However, a gaseous eccentric disc may be unstable to hydrodynamic instabilities, and these may limit the lifetime of the eccentricity.

In a companion paper (\citealt{OB2014a}; hereafter OB14), we derived a local model of an eccentric Keplerian disc, which can be used to study its linear stability, as well as for future nonlinear numerical studies. We showed how the dynamics of the local model can be used to determine the evolution of the mass, angular momentum and eccentricity distributions in the disc. We also derived the nonlinear vertical oscillations of the disc (first obtained in a global model by \citealt{Ogilvie2001}) and studied their behaviour numerically. In an isothermal disc the vertical oscillations exhibit extreme behaviour for eccentricities above approximately $0.5$, which could potentially lead to shocks and resulting dissipation in their nonlinear evolution. Such discs are likely to evolve violently on a dynamical timescale, so we concentrate instead on discs with smaller eccentricities. In this case, the behaviour of the vertical oscillations is a regular ``breathing mode" of the disc, which causes an additional periodic variation of the fluid properties around an orbit. Such discs may be unstable to parametric instabilities involving low-frequency internal waves. These may lead to a damping of the disc eccentricity and could limit the eccentricities of observable discs. In this paper we use the local model to study the linear stability of an eccentric disc.

\cite{John2005a} was the first to study the hydrodynamic instability of eccentric Keplerian discs, motivated by the earlier work of \cite{Goodman1993} and \cite{LubowPringleKerswell1993} for tidally deformed discs (which have $m=2$). He found that eccentric discs were unstable to a parametric instability that took the form of resonantly excited inertial waves. Their nonlinear evolution in a global disc model was subsequently studied in \cite{John2005b}, where they led to small-scale subsonic turbulence (or wave activity) and to a gradual decay of the disc eccentricity. The pioneering calculations of \cite{John2005a} were limited to studying the three-dimensional stability of a uniformly eccentric disc without vertical structure. In this paper we use the newly derived local model to study the local linear stability of discs with any eccentricity and eccentricity gradient, taking into account the vertical structure of the disc in full.

The structure of this paper is as follows. In \S \ref{basiceqns} we write down the equations describing fluid dynamics within the local model derived in OB14, and describe the resulting vertical oscillatory flows in the disc. We set up the local linear stability analysis of an eccentric disc in \S \ref{linstab}, analyse this system numerically in \S \ref{Numerical}, and end with a discussion and conclusion. Detailed analytical understanding of the instability is relegated to Appendices \ref{Theory} to \ref{WKBTheory}.

\section{Local model and laminar flows}
\label{basiceqns}
The basic equations describing ideal isothermal hydrodynamics within the local model of a coplanar eccentric Keplerian disc are summarised in this section. The properties of an eccentric orbit in a coplanar Keplerian disc can be described by the following parameters: the semi-latus rectum $\lambda$ (related to the semi-major axis $a$ by $\lambda=a(1-e^2)$), the eccentricity $e(\lambda)$, and the longitude of pericentre $\omega(\lambda)$. When formulated in dimensionless terms, the local model is independent of $\lambda$, but does depend on the local eccentricity $e$, as well as the dimensionless local gradients in the eccentricity $\lambda e^{\prime}\equiv\lambda \mathrm{d}e/\mathrm{d}\lambda$ and $e \lambda \omega^{\prime}\equiv e \lambda \mathrm{d}\omega/\mathrm{d}\lambda$, where the latter may be thought of as a measure of the twist in the disc. These may be combined into the complex eccentricity $E=e \mathrm{e}^{\mathrm{i}\omega}$ and eccentricity gradient $\lambda E^{\prime}=\lambda \mathrm{d}E/\mathrm{d}\lambda$.

The local model derived in OB14 is valid for a thin disc with $\epsilon = H/r \ll 1$ and describes fluid dynamics in a small patch of the disc centred around a reference orbit at the mid-plane with orbital coordinates $(\lambda_{0},\varphi (t),0)$.
Owing to the geometry of an eccentric orbit, it is convenient to adopt (in general) non-orthogonal coordinates ($\xi,\eta,\zeta$), where $\xi=\lambda-\lambda_0$ is a quasi-radial coordinate, $\eta=\phi-\varphi (t)$ is an angular coordinate and $\zeta=z$ is the usual vertical coordinate. The coordinates $(\xi,\lambda_{0}\eta,\zeta)$ are equivalent to Cartesian coordinates when the orbit is circular -- in this case the system of equations that we will list below reduces to the standard shearing box commonly used to study the dynamics of astrophysical discs. 

We define the contravariant velocity components $(v^{\xi},v^{\eta},v^{\zeta})$ and the enthalpy $h$, where the latter is defined by
\begin{eqnarray}
h=c_{s}^{2}\ln \rho + \mathrm{const},
\end{eqnarray} 
for an isothermal ideal gas with sound speed $c_{s}$, in which the pressure $p$ is related to the density $\rho$ by $p=c_{s}^{2}\rho$. 

The linear stability of an eccentric disc to a general non-axisymmetric disturbance is complicated considerably by the presence of Keplerian shear, so we consider locally axisymmetric motions in this work. The resulting (inviscid) equations in the local model are (Eq.~80--83 in OB14)
\begin{eqnarray}
\label{Eq1}
&& \mathrm{D}v^{\xi}+2\Gamma^{\lambda}_{\lambda \phi} \Omega v^{\xi}+2\Gamma^{\lambda}_{\phi\phi} \Omega v^{\eta}=-g^{\lambda\lambda}\partial_{\xi}h, \\
&& \mathrm{D}v^{\eta}+(\partial_{\lambda}\Omega+2\Gamma^{\phi}_{\lambda \phi} \Omega) v^{\xi}+(\partial_{\phi}\Omega+2\Gamma^{\phi}_{\phi\phi}\Omega) v^{\eta}= \\
\nonumber
&& \hspace{2.2in} -g^{\lambda\phi}\partial_{\xi}h , \\
&&  \mathrm{D}v^{\zeta}=-\Phi_{2}\zeta -\partial_{\zeta}h, \\
&&  \mathrm{D}h=-c_{s}^{2}\left(\Delta +\partial_{\xi}v^{\xi} +\partial_{\zeta}v^{\zeta}\right),
\label{Eq4}
\end{eqnarray}
where
\begin{eqnarray}
\mathrm{D}\equiv \partial_{t} +v^{\xi}\partial_{\xi}+v^{\zeta}\partial_{\zeta},
\end{eqnarray}
is the Lagrangian derivative. We have evaluated the orbital angular velocity $\Omega$ and its derivatives, as well as the metric and connection coefficients, at a reference point in the mid-plane of the disc $(\lambda_{0},\varphi (t),0)$, so that these become periodic functions of time only. The orbital velocity divergence is written as $\Delta$, which is nonzero when the disc has an eccentricity gradient (see OB14 and Appendix \ref{Coefficients}).

The gravitational potential expanded about the mid-plane takes the form (with $\zeta=O(\epsilon)$)
\begin{eqnarray}
\Phi=\Phi_{0}+\frac{1}{2}\zeta^{2}\Phi_{2}+O(\zeta^4),
\end{eqnarray}
where $\Phi_{0}=-GM/R$ and $\Phi_{2}=GM/R^3$, and $R$ is the cylindrical radius. The periodic variation of $\Phi_{2}$ around an orbit is responsible for driving vertical oscillatory flows in the disc. These (nonlinear) oscillations can be obtained by looking for simple solutions of Eqs.~\ref{Eq1}--\ref{Eq4} of the form
\begin{eqnarray}
v^{\xi}=v^{\eta}=0, \;\;\;\;\;\; v^{\zeta}=w(t)\zeta, \;\;\;\;\;\; h=f(t)-\frac{1}{2}\zeta^{2} g(t), 
\end{eqnarray}
which satisfy the following ODEs:
\begin{eqnarray}
\label{laminareqns}
&& \mathrm{d}_{t}w+w^2=-\Phi_{2}+g, \\
&& \mathrm{d}_{t}f=-c_{s}^{2}\left(\Delta+w\right), \\
&& \mathrm{d}_{t}g=-2wg.
\label{laminareqns2}
\end{eqnarray}
The laminar flow functions $f$ and $g=c_{s}^{2} H^{-2}$ (where $H(t)$ is the Gaussian scaleheight of the isothermal disc), together define the surface density of the disc
\begin{eqnarray}
\Sigma\propto \mathrm{e}^{\frac{f}{c_{s}^{2}}}c_{s}g^{-\frac{1}{2}}\propto\mathrm{e}^{\frac{f}{c_{s}^{2}}}H,
\end{eqnarray}
satisfying
\begin{eqnarray}
\mathrm{d}_{t}\Sigma=-\Delta \Sigma,
\end{eqnarray}
so that the surface density is constant around an elliptical orbit when $\Delta=0$, but varies if $\Delta\ne 0$. 

Periodic solutions of Eqs.~\ref{laminareqns}--\ref{laminareqns2} can be computed numerically using a shooting method. Several examples have been plotted in OB14. Note that $\lambda e^{\prime}$ and $e \lambda \omega^{\prime}$ play no role in determining $g$ and $w$ in the isothermal approximation (although the enthalpy at the mid-plane $f$ does depend on $\Delta$).

\section{Linear stability of eccentric discs}
\label{linstab}
\subsection{Linearised axisymmetric perturbation equations}
We consider small perturbations to the orbital motion and vertical laminar flows of the form $\mathrm{Re}\left[\hat{v}^{\xi}(\zeta,t)\mathrm{e}^{\mathrm{i} k_{\xi}\xi} \right]$, and so on for other variables, where $k_{\xi}$ is a quasi-radial wavenumber. We subsequently drop the hat on the perturbations for clarity. We choose units such that $\Omega_{0}=\left(\frac{GM}{\lambda_{0}^3}\right)^{\frac{1}{2}}=1$ and $c_{s}=1$, therefore the disc thickness would take a constant value $H=g^{-\frac{1}{2}}=1$ for a circular disc.

We note that it is much simpler to use the true anomaly $\theta$ as a variable rather than $t$, and that the corresponding rates of change are related by $\partial_{t}=\Omega\partial_{\theta}=(1+e \cos\theta)^{2}\partial_{\theta}$. This variable is used as a ``time-like" variable, and is continuous and monotonically increasing, not restricted to the range $[0,2\pi]$. We further define $c\equiv\cos\theta$ and $s\equiv \sin \theta$. The resulting linearised perturbation equations are
\begin{eqnarray}
\label{Eq1a}
&& \hspace{-0.5cm} \Omega\partial_{\theta}v^{\xi}+ w \zeta \partial_{\zeta} v^{\xi}+2\Gamma^{\lambda}_{\lambda \phi} \Omega v^{\xi}+2\Gamma^{\lambda}_{\phi\phi}\Omega v^{\eta}=-\mathrm{i} g^{\lambda\lambda} k_{\xi}h, \\
\nonumber
&& \hspace{-0.5cm} \Omega\partial_{\theta}v^{\eta}+ w \zeta \partial_{\zeta} v^{\eta} +\left(\partial_{\lambda}\Omega+ \Gamma^{\phi}_{\lambda \phi} \Omega\right)v^{\xi}\\
&& \hspace{0.9in} +\left(\partial_{\phi}\Omega+2\Gamma^{\phi}_{\phi\phi}\right)v^{\eta} =-\mathrm{i} g^{\lambda\phi}k_{\xi}h, \\
&& \hspace{-0.5cm} \Omega\partial_{\theta}v^{\zeta}+ w \zeta \partial_{\zeta} v^{\zeta}+w v^{\zeta}=-\partial_{\zeta}h, \\
&& \hspace{-0.5cm} \Omega\partial_{\theta}h+ w \zeta \partial_{\zeta} h-g \zeta v^{\zeta}=-\mathrm{i} k_{\xi} v^{\xi}-\partial_{\zeta}v^{\zeta}.
\label{Eq4a}
\end{eqnarray}
The coefficients are periodic functions of $\theta$ with period $2\pi$, which we list in Appendix \ref{Coefficients}. A statement of conservation of energy for the perturbations can be derived at this stage, which we present in Appendix \ref{Energy} -- this is used to analyse the energetics of the resulting instabilities.

The vertical structure of waves in a circular isothermal disc take the form of Hermite polynomials in $\zeta$ (e.g.~\citealt{Okazaki1987,OL2013b}). These have the property that the energy density of the perturbations tends to zero at large distances from the mid-plane. In the eccentric case, Eqs.~\ref{Eq1a}--\ref{Eq4a} also have exact solutions that are polynomials in $\zeta$. These solutions can be represented as finite sums of (probabilist's) Hermite polynomials:
\begin{eqnarray}
v^{\xi}&=&\sum_{n=0}^{N}u^{\xi}_{n}(\theta)\mathrm{He}_{n}(\zeta), \\
v^{\eta}&=&\sum_{n=0}^{N}\lambda^{-1} u^{\eta}_{n}(\theta)\mathrm{He}_{n}(\zeta), \\
v^{\zeta}&=&\sum_{n=1}^{N}u^{\zeta}_{n}(\theta)\mathrm{He}_{n-1}(\zeta), \\
h&=&\sum_{n=0}^{N}h_{n}(\theta)\mathrm{He}_{n}(\zeta),
\end{eqnarray}
where $N$ is the vertical mode number. The factor of $\lambda^{-1}$ ensures that $u^{\eta}_{n}$ has units of velocity.
Note that
\begin{eqnarray}
\mathrm{d}_{\zeta} \mathrm{He}_{n}(\zeta)&=&n\mathrm{He}_{n-1}(\zeta), \\
\zeta \mathrm{He}_{n}(\zeta)&=&\mathrm{He}_{n+1}(\zeta)+n \mathrm{He}_{n-1}(\zeta).
\end{eqnarray}
The resulting ODEs are
\begin{eqnarray}
\label{laminar}
&&  (1+e c )^{2}\mathrm{d}_{\theta}w+w^2=-\Phi_{2}+g, \\
&&  (1+e c)^{2}\mathrm{d}_{\theta}f=-\left(\Delta+w\right), \\
&&  (1+e c)^{2}\mathrm{d}_{\theta}g=-2wg,
\label{laminar1}
\end{eqnarray}
for the laminar flows, and
\begin{eqnarray}
\label{EqH1}
&& (1+ec)^{2}\mathrm{d}_{\theta}u^{\xi}_{n}+w\left[n u^{\xi}_{n}+(n+1)(n+2)u^{\xi}_{n+2}\right] \\
\nonumber 
&& \hspace{0.1in} +2\Gamma^{\lambda}_{\lambda \phi} \Omega u^{\xi}_{n}+2\lambda^{-1}\Gamma^{\lambda}_{\phi\phi} \Omega u^{\eta}_{n}=-\mathrm{i} g^{\lambda\lambda}k_{\xi}h_{n}, \\
\label{EqH2}
&& (1+ec)^{2}\mathrm{d}_{\theta}u^{\eta}_{n}+w\left[n u^{\eta}_{n}+(n+1)(n+2)u^{\eta}_{n+2}\right]\\ 
\nonumber
&& \hspace{0.1in} +u_{n}^{\xi}\lambda \partial_{\lambda}\Omega + 2\lambda\Gamma^{\phi}_{\lambda \phi} \Omega u^{\xi}_{n}+\Gamma^{\phi}_{\phi\phi} \Omega u^{\eta}_{n}=-\mathrm{i} \lambda g^{\lambda\phi}k_{\xi}h_{n}, \\
\label{EqH3}
&& (1+ec)^{2}\mathrm{d}_{\theta}u^{\zeta}_{n}+w\left[(n-1) u^{\zeta}_{n}+n(n+1)u^{\zeta}_{n+2}\right] \\
\nonumber
&& \hspace{0.1in} + w u^{\zeta}_{n}=-n h_{n}, \\
\label{EqH4}
&& (1+ec)^{2}\mathrm{d}_{\theta}h_{n}+w\left[n h_{n}+(n+1)(n+2)h_{n+2}\right]\\
\nonumber
&& \hspace{0.1in} -(g-1)\left[u^{\zeta}_{n}+(n+1)u^{\zeta}_{n+2}\right]=-\mathrm{i}k_{\xi}u^{\xi}_{n}+u^{\zeta}_{n},
\end{eqnarray}
for the linear perturbations. Eq.~\ref{EqH3} is valid for $1 \leq n \leq N$ and the other three are valid for $0 \leq n \leq N$. We have set $u^{\xi}_{n}=0$ etc for $n>N$ by considering polynomial solutions. This is the system of equations that we will solve to determine the linear stability of an eccentric disc to locally axisymmetric perturbations. Note that vertical hydrostatic equilibrium corresponds to $g=H^{-2}=1$, which does not hold in an eccentric disc, in general. Also, note that there are three dimensionless parameters that define the local orbital properties of the eccentric disc: $e, \lambda e^{\prime}, e\lambda \omega^{\prime}$.

For a given vertical mode number $N$, the set of $4N-1$ equations given by Eqs.~\ref{EqH1}--\ref{EqH4} are analysed numerically using a Floquet method. This method is appropriate since the coefficients are periodic functions of $\theta$. First, the monodromy matrix of linearly independent solutions is constructed by integrating the ODEs over one period (in the process the laminar flows are also computed) for initial conditions such that all variables except one are set to zero. The eigenvalues of the monodromy matrix allow us to obtain the complex growth rates of the instability. See \cite{OL2013b} for details of a similar approach used to study the instabilities of a warped disc. We have verified that we obtain the correct linear dispersion relation for a circular disc, and we will illustrate in \S \ref{Numerical} that our numerically computed growth rates are in excellent agreement with the analytical predictions presented in Appendix \ref{Theory}. 

The vertical oscillations of the disc couple components with different $n$ when $e\ne 0$. However, these only couple a component $n$ with a component $m=n+2$, and there are no additional couplings to components with $m<n$. These are then ``one-way" couplings, for which components with $m<n$ are slaved to the maximum $n$. The growth rate of the instability observed in the next section is therefore fully determined by considering only the component with $n=N$.

\subsection{Linear axisymmetric waves in a circular disc}
When $e=\lambda e^{\prime}=e\lambda\omega^{\prime}=w=g-1=0$, the above system reduces to
\begin{eqnarray}
&& \mathrm{d}_{\theta}u^{\xi}_{n} -2\Omega u^{\eta}_{n}=-\mathrm{i}k_{\xi}h_{n}, \\
&& \mathrm{d}_{\theta}u^{\eta}_{n}+\frac{1}{2}\Omega u^{\xi}_{n}=0, \\
&& \mathrm{d}_{\theta}u^{\zeta}_{n}=-n h_{n}, \\
&& \mathrm{d}_{\theta}h_{n}=-\mathrm{i} k_{\xi}u^{\xi}_{n}+u^{\zeta}_{n}.
\end{eqnarray}
Looking for solutions proportional to $\mathrm{e}^{-\mathrm{i}\omega\theta}$, we obtain the ideal dispersion relation describing axisymmetric waves in a circular isothermal Keplerian disc:
\begin{eqnarray}
(-\omega^{2}+n)(-\omega^{2}+1)-k_{\xi}^{2}\omega^{2}=0.
\end{eqnarray}
The low frequency branch for $n\ne 0$ corresponds to inertial waves. A pair of inertial waves with $\omega=\pm \frac{1}{2}$ (which occurs when $k_{\xi}=\frac{1}{2}\sqrt{3(4n-1)}$) can be coupled to give $1$, which is the frequency at which the geometrical coefficients in Eqs.~\ref{EqH1}--\ref{EqH4} are modulated. For a given range of $k_{\xi}$ these resonant waves are fully captured by considering a finite range of vertical mode numbers. These are the waves that become unstable in an eccentric disc, as we will illustrate in the next section.

\section{Numerical calculations}
\label{Numerical}
In a disc with nonzero eccentricity or eccentricity gradient the coefficients in Eqs.~\ref{EqH1}--\ref{EqH4} become $2\pi$-periodic functions of $\theta$. The periodic variation in the eccentric orbital motion of the gas around an orbit, together with the associated vertical oscillation of the disc, drive a parametric instability consisting of pairs of inertial waves. In the local model, the eccentric orbital motion has radial and vertical wavenumbers of 0 and a frequency of 1 (i.e.~the orbital frequency). For small departures from circularity, an instability is driven by a parametric resonance between the eccentric (and vertical) oscillation of the fluid around an orbit and a pair of inertial waves with $\omega=\pm1/2$ with the same radial and vertical wavenumbers. These waves propagate radially in opposite directions, and their superposition is a standing wave. These are coupled in an eccentric disc because their frequencies differ by 1. When the departure from circularity is not small (or if viscosity is included), there is a frequency band of instability around exact resonance whose width increases with the eccentricity. The instability of a disc with an eccentricity gradient is found to take the same form, as we will illustrate below. This instability is explained in detail in Appendix \ref{Theory}, where we analytically compute the growth rates at exact resonance. We have also analysed the sources of energy driving the instability, which we present in Appendix \ref{Energy}.

\subsection{Uniformly eccentric disc}
\label{uniforme}
\begin{figure*}
  \begin{center}
    \subfigure[$e=0.01$]{\includegraphics[trim=6cm 0cm 6cm 0cm, clip=true,width=0.485\textwidth]{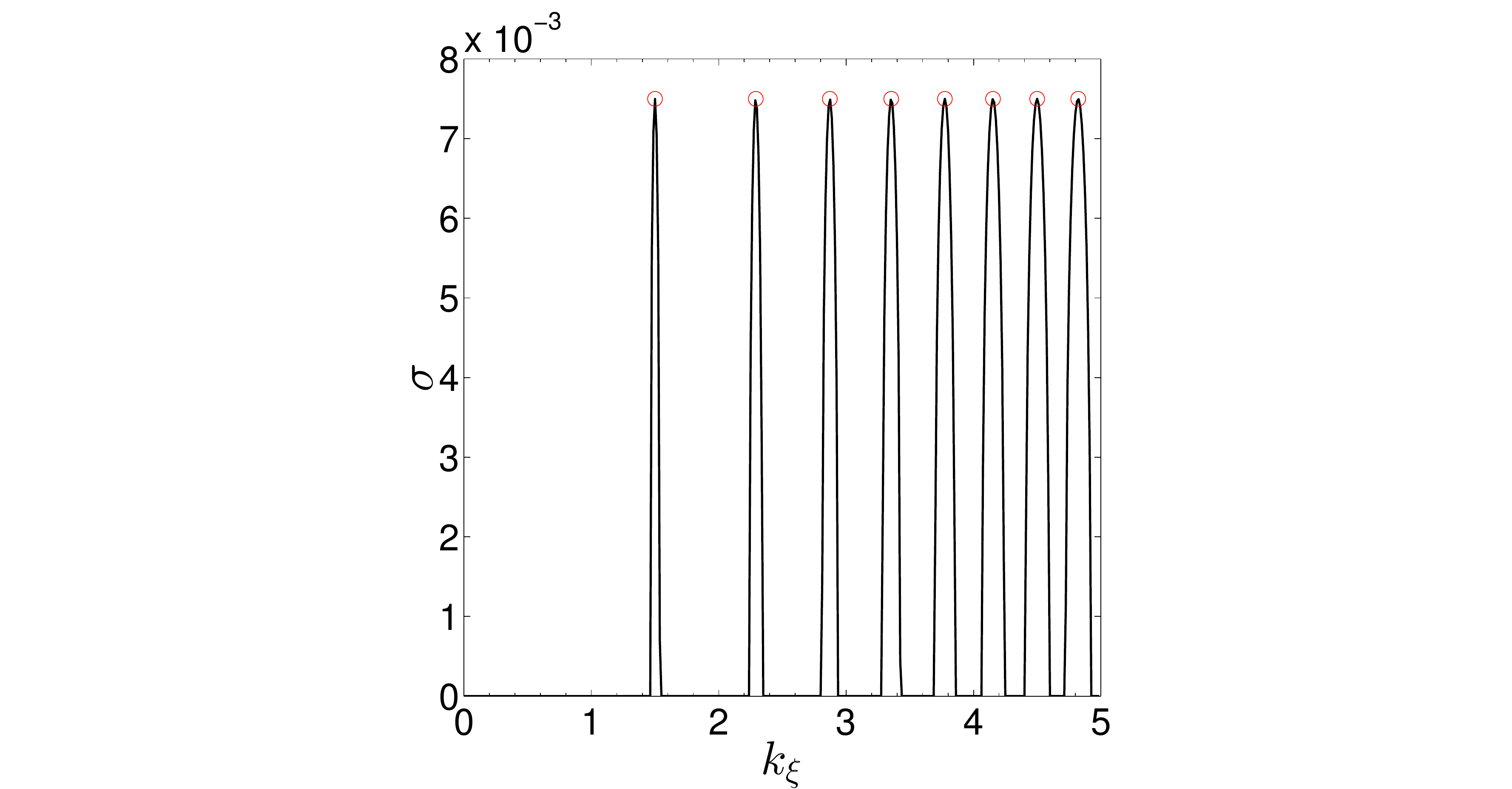} }
    \subfigure[$e=0.05$]{\includegraphics[trim=6cm 0cm 6cm 0cm, clip=true,width=0.485\textwidth]{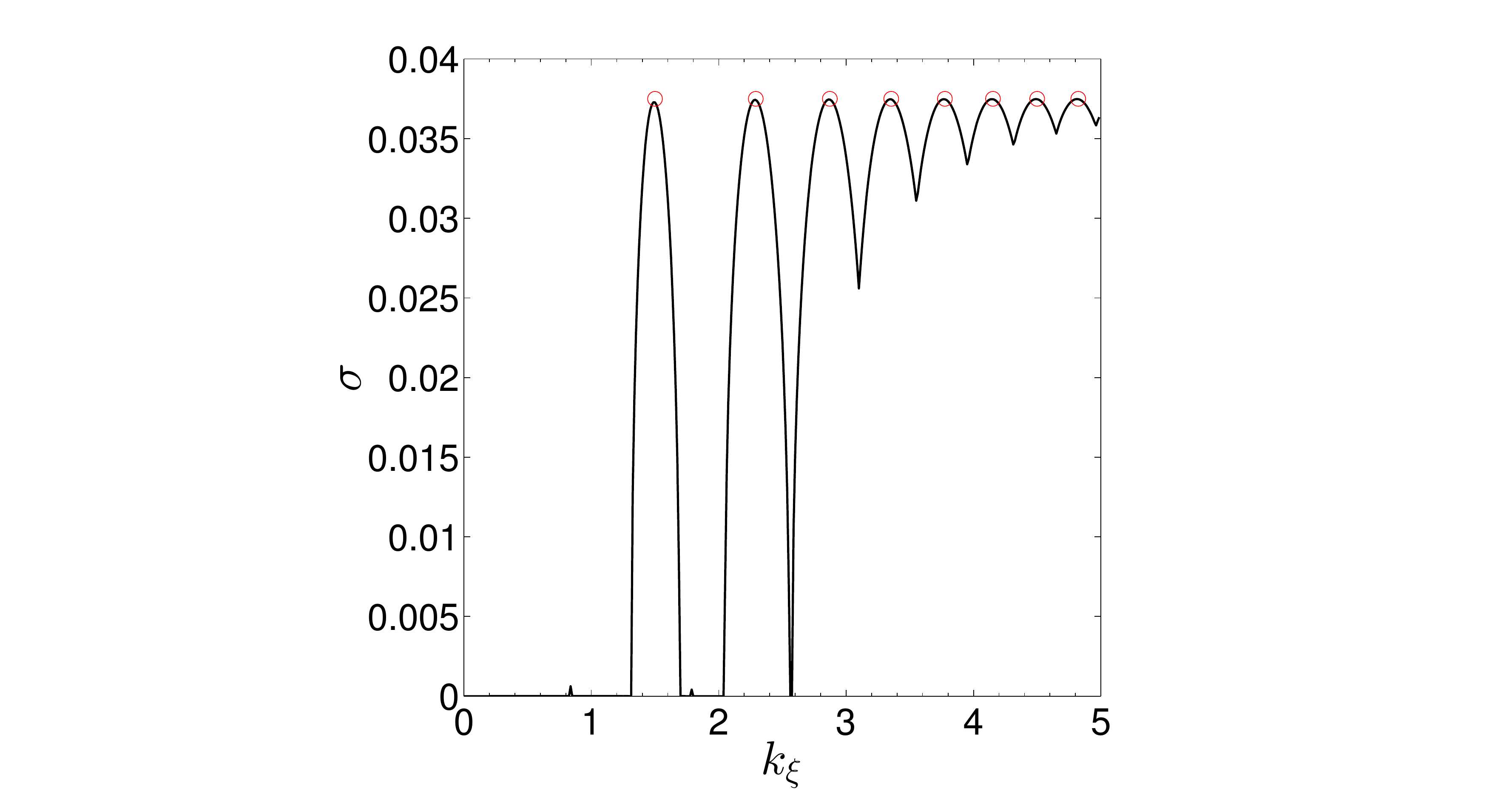} } 
    \subfigure[$e=0.1$]{\includegraphics[trim=6cm 0cm 6cm 0cm, clip=true,width=0.485\textwidth]{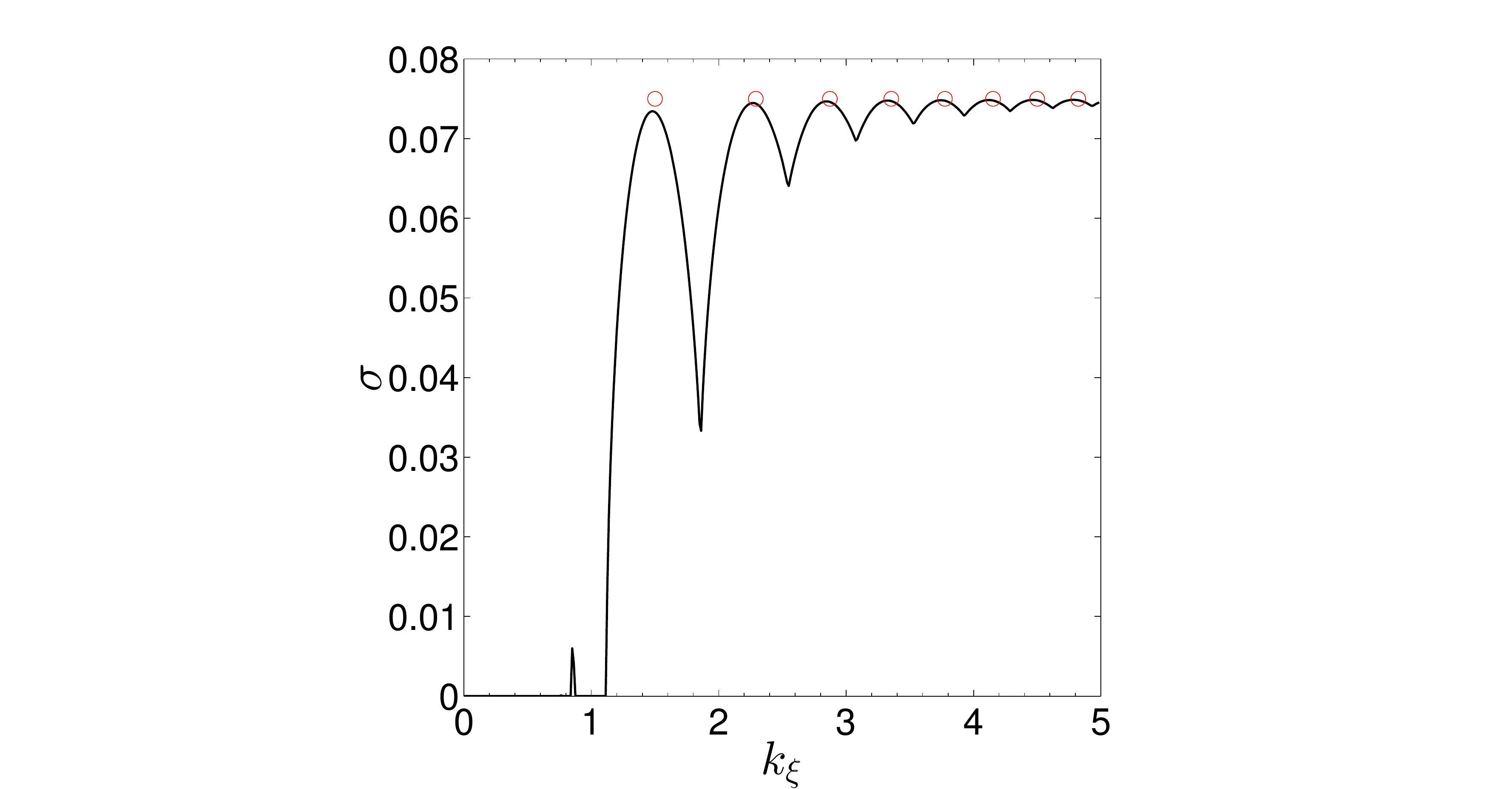} } 
    \subfigure[$e=0.2$]{\includegraphics[trim=6cm 0cm 6cm 0cm, clip=true,width=0.485\textwidth]{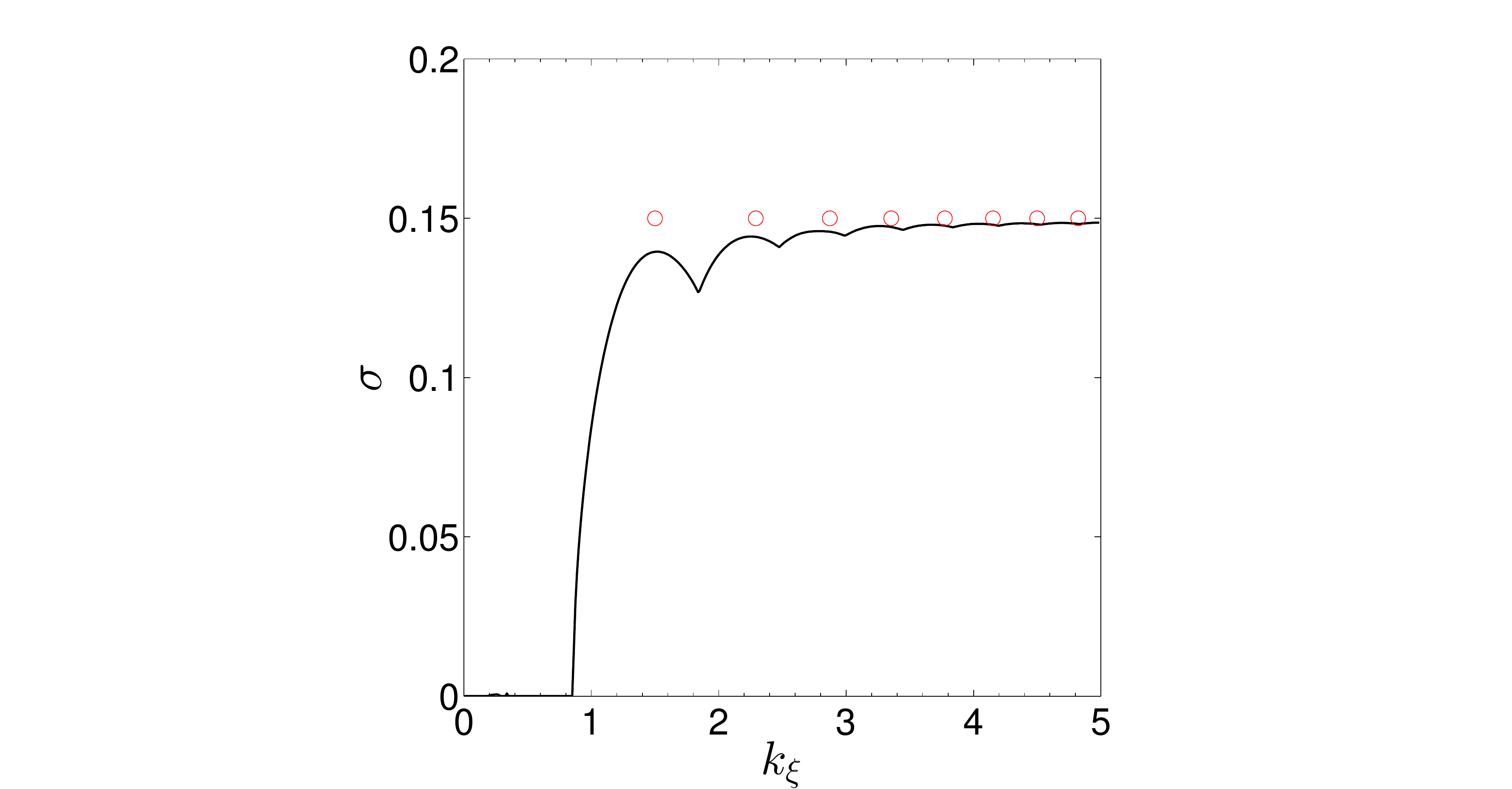} }  
    \subfigure[$e=0.4$]{\includegraphics[trim=6cm 0cm 6cm 0cm, clip=true,width=0.485\textwidth]{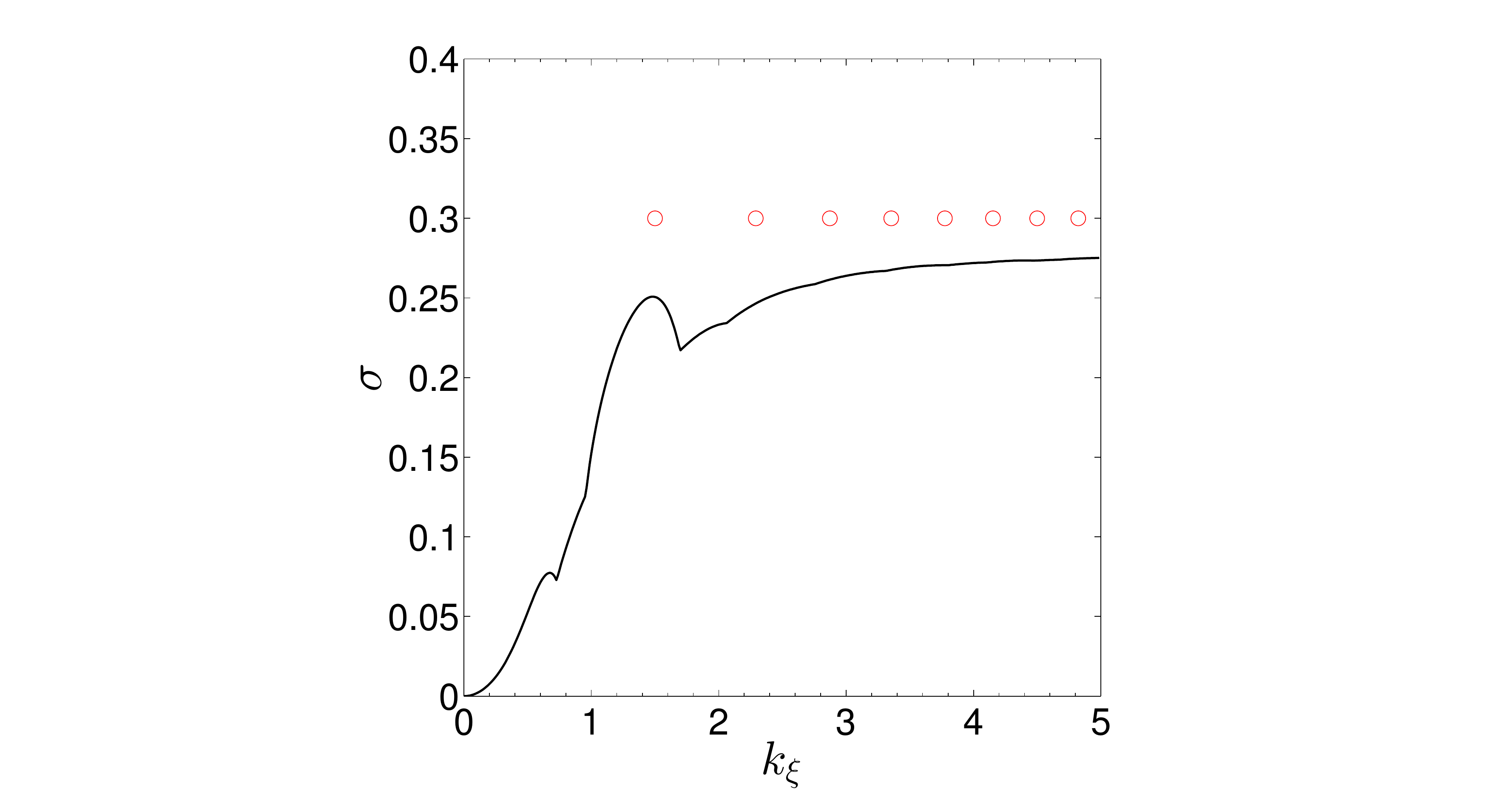} } 
    \subfigure[$e=0.55$]{\includegraphics[trim=6cm 0cm 6cm 0cm, clip=true,width=0.485\textwidth]{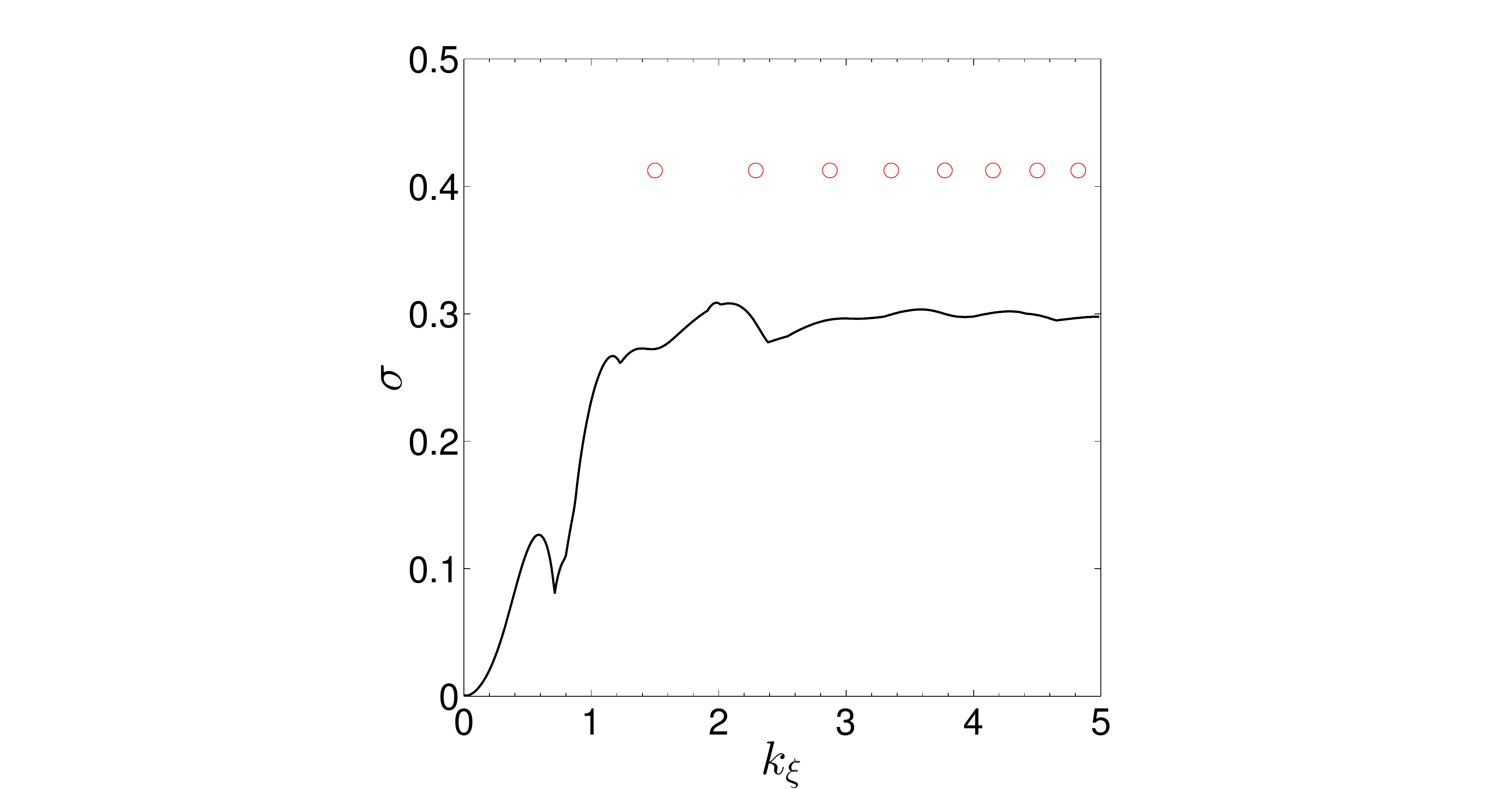} } 
    \end{center}
  \caption{Instability of a uniformly eccentric disc with $\lambda e^{\prime}=\lambda \omega^{\prime}=0$. The growth rate of the fastest growing mode (using units such that $\left(GM/\lambda_{0}^3\right)^{\frac{1}{2}}=1$) is plotted as a function of $k_{\xi}$ for various $e$. The red circles show the analytical prediction from Appendix \ref{Theory} at exact resonance, which is valid when $e\ll 1$. The numerical calculations were performed with vertical mode numbers up to $N=12$, which was sufficient to obtain the fastest growing mode in all cases, and involved calculations at $400$ uniformly distributed values of $k_{\xi}\in[0,5]$. Computations for larger $e$ were not performed owing to the extreme behaviour of the vertical laminar flows in these cases.}
  \label{1}
\end{figure*}

We first illustrate the instability of a uniformly eccentric disc, with $\lambda e^{\prime}=\lambda \omega^{\prime}=0$.  In Fig.~\ref{1}, we plot the growth rate of the fastest growing mode from our numerical calculations as a function of the radial wavenumber $k_{\xi}$ for various $e$. We have also plotted our analytical predictions from Appendix \ref{Theory} for the growth rate ($\sigma$) at exact resonance as red circles, where $\sigma=3 e/4$ independent of $k_{\xi}$ when $e\ll1$.  For small $e$, instability occurs in discrete wavenumber bands centred on certain values of $k_{\xi}$, which merge as $k_{\xi}\rightarrow \infty$. The first peak represents a pair of inertial waves with $n=1$, therefore $k_{\xi}=3/2$. The subsequent peaks represent inertial waves with sequentially increasing $n$ in such a way that the waves have $\omega=\pm \frac{1}{2}$. As $e$ is increased, the instability bands become wider and merge, and their centres are shifted slightly from the analytical prediction. There is an additional peak at small $k_{\xi}$ below the first instability band whose growth rate is $O(e^2)$; this instability is also found to have an inertial wave character. For $e\gtrsim 0.4$, there is instability for any $k_{\xi}>0$. 

\begin{figure*}
  \begin{center}
         \subfigure[$\theta=0$]{\includegraphics[trim=6cm 0cm 6cm 0cm, clip=true,width=0.48\textwidth]{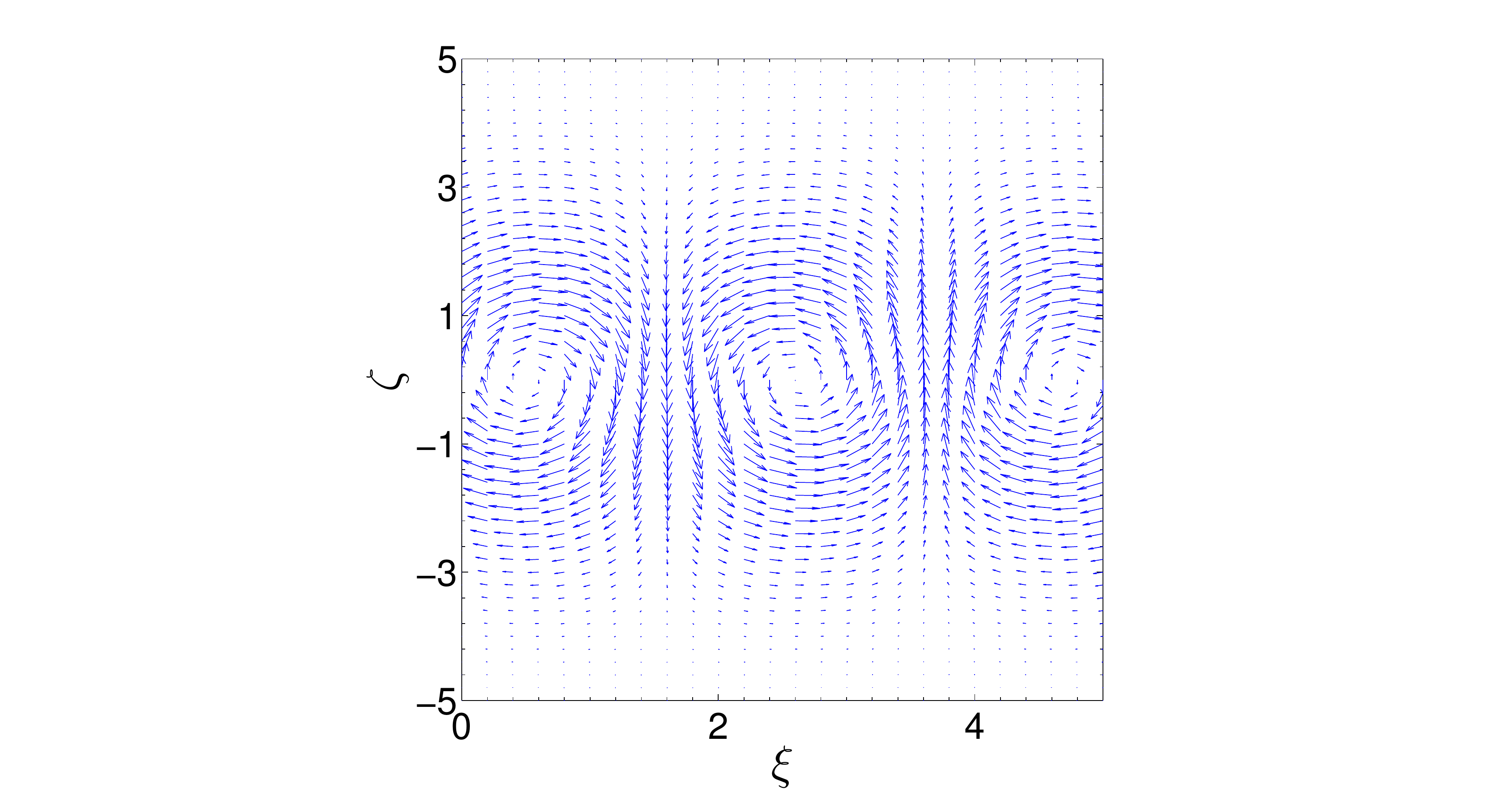} }
      \subfigure[$\theta=\pi/2$]{\includegraphics[trim=6cm 0cm 6cm 0cm, clip=true,width=0.48\textwidth]{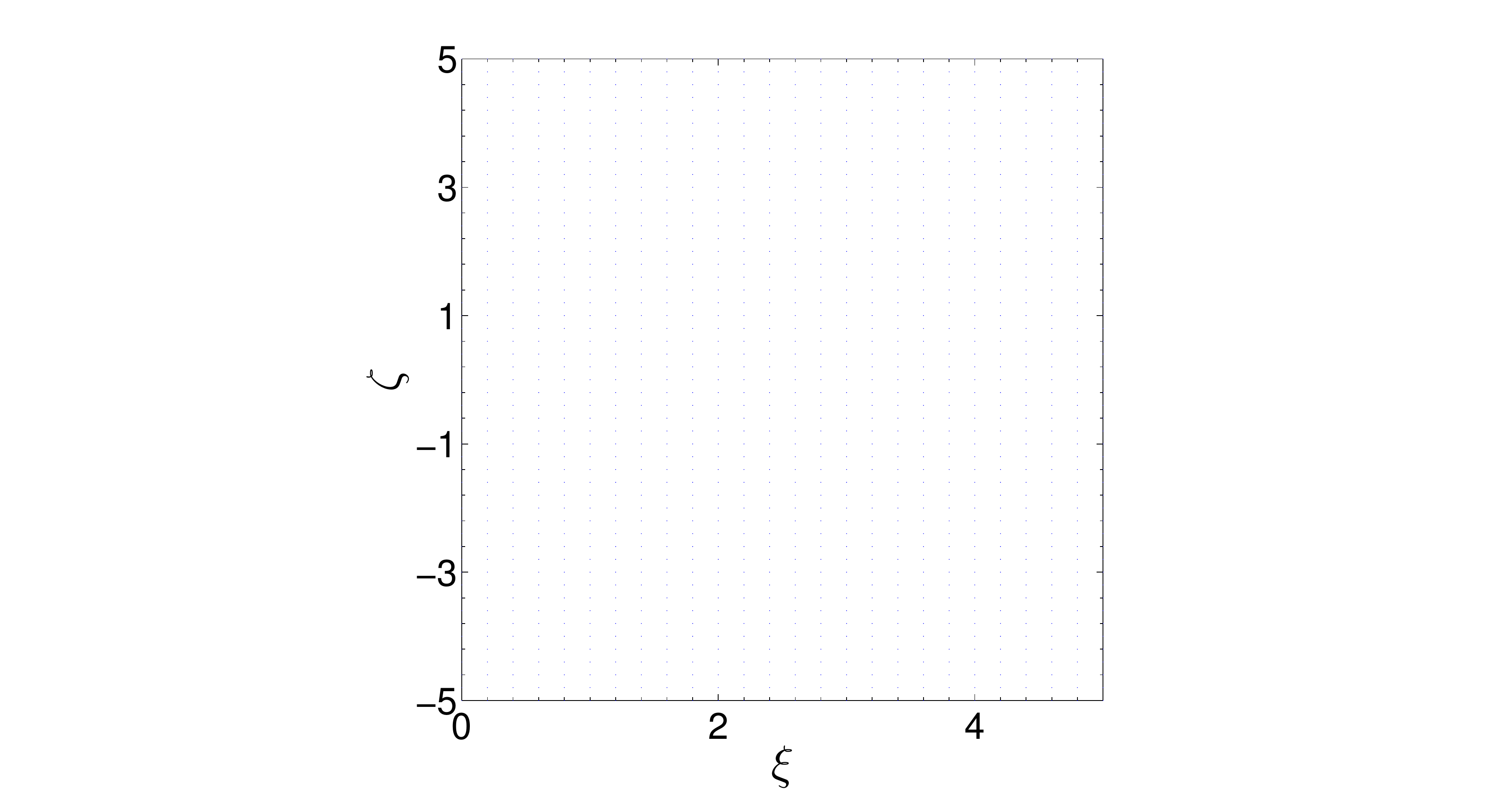} } \\
       \subfigure[$\theta=\pi$]{\includegraphics[trim=6cm 0cm 6cm 0cm, clip=true,width=0.48\textwidth]{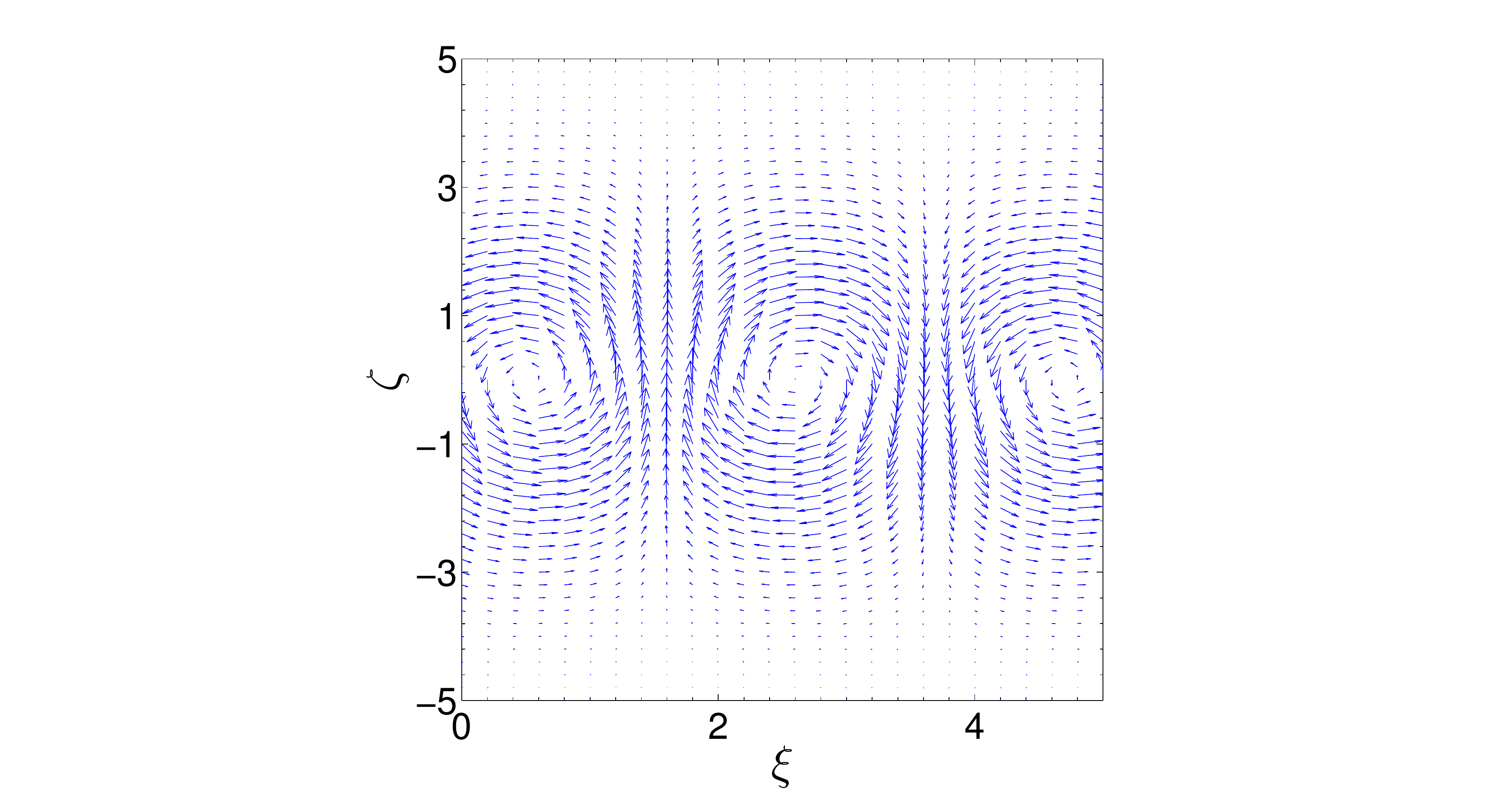} }
        \subfigure[$\theta=3\pi/2$]{\includegraphics[trim=6cm 0cm 6cm 0cm, clip=true,width=0.48\textwidth]{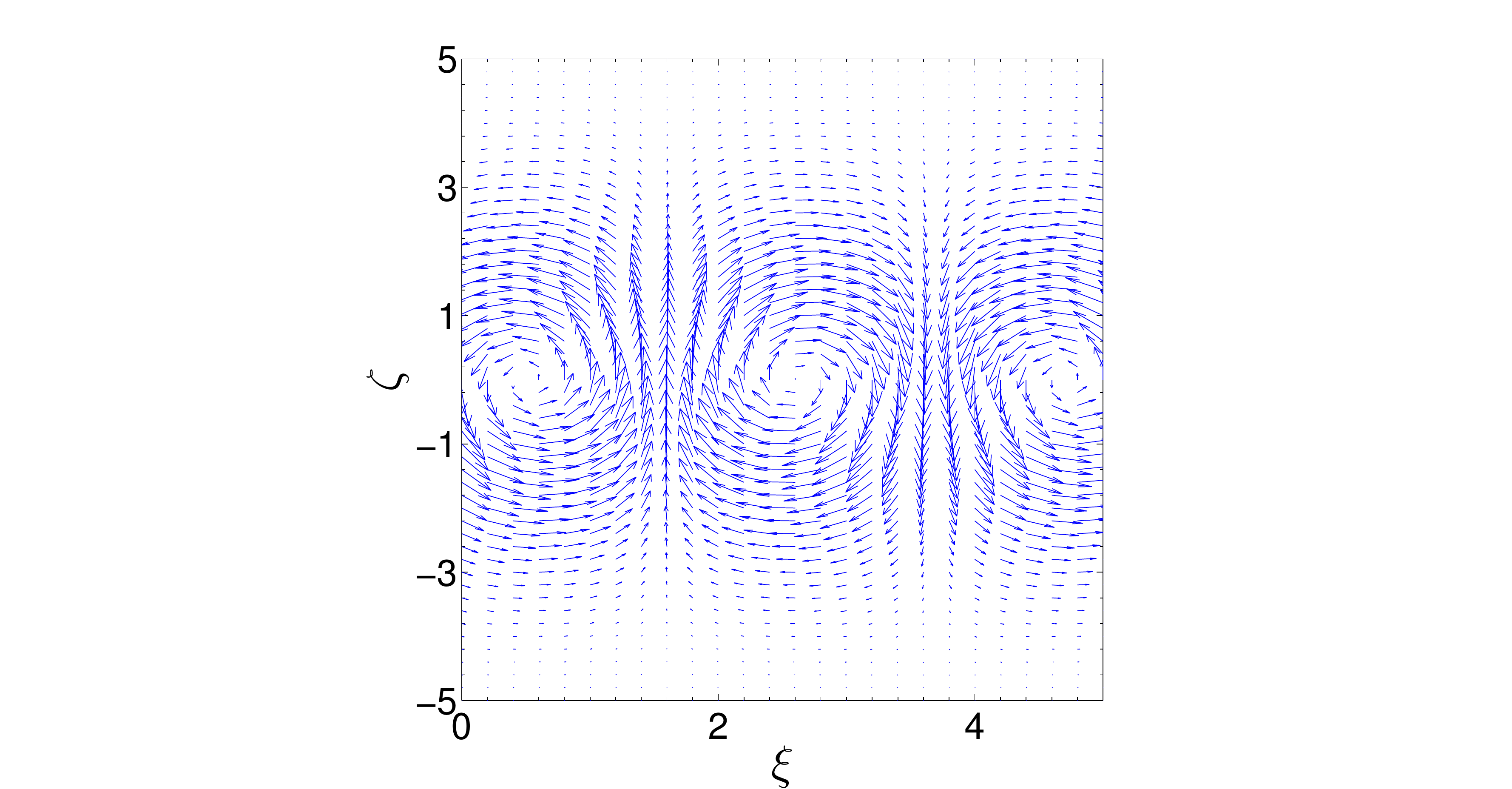} }
          \end{center}
          \caption{Illustration of the velocity field (multiplied by $\sqrt{\rho}$ to show the localisation of wave energy near to the mid-plane) of the $n=1$ unstable mode at exact resonance when $e=0.01$ for four different phases around an orbit. This is a standing mode composed of a superposition of travelling inertial waves propagating in opposite directions radially. The arrows have the same scale in each panel. When the ``meridional" velocity perturbations vanish at $\theta=\frac{\pi}{2}$, the azimuthal velocity perturbation is nonzero. The amplitude of the vertical velocity is correlated with the orbital motion in such a way to extract energy from the vertical oscillation of the disc, and is maximum at $\theta=\frac{3\pi}{2}$, where $-w$ has its maximum. At $\theta=2\pi$, the mode returns to its form at $\theta=0$ but is slightly amplified and reversed in sign since its period is $4\pi$.}
  \label{2}
\end{figure*}

We illustrate the velocity field for one representative unstable mode when $e=0.01$ in the $(\xi,\zeta)$-plane in Fig.~\ref{2} -- the velocity components $u^{\xi}$ and $u^{\zeta}$ have been multiplied by $\rho^{1/2}$ to show the wave energy at four different phases around an orbit. This mode is a standing wave, whose amplitude is modulated in such a way to extract energy from the orbital flow and the vertical oscillation of the disc. The dominant contribution comes from the vertical oscillation of the disc, arising from a term $\mathrm{Re}\left[-w|u^{\zeta}|^{2}\right]$ in the energy equation, which has net contribution $\propto -\int_{0}^{2\pi} \sin\theta (2-2\sin \theta) d\theta = 2\pi$ (this is shown in detail in Appendix \ref{Energy}). The mode has its maximum magnitude of vertical velocity at the phase $\theta=3\pi/2$, at which $-w$ has its maximum, therefore it can extract energy from the vertical oscillation most efficiently at this phase.

\begin{figure}
  \begin{center}
    \subfigure{\includegraphics[trim=6cm 0cm 6cm 0cm, clip=true,width=0.49\textwidth]{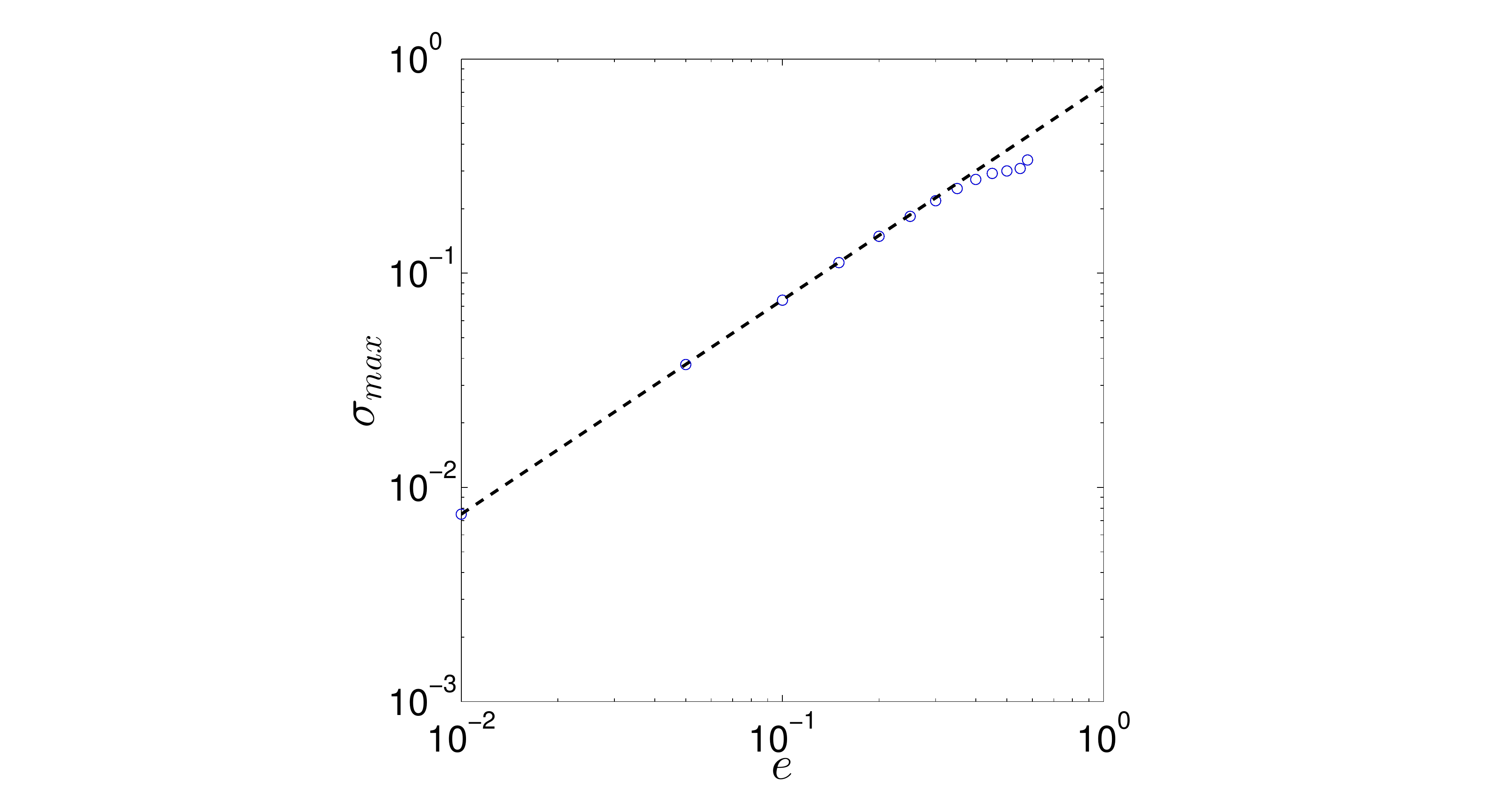} }
    \end{center}
  \caption{Maximum growth rate of the instability (for $k_{\xi}\in [0,5]$) as a function of $e$ for a uniformly eccentric disc (using units such that $\left(GM/\lambda_{0}^3\right)^{\frac{1}{2}}=1$). The numerical results are plotted as blue circles, and were computed up to $N=12$, which was found to be sufficient to obtain the fastest growing mode in all cases. The black dashed line is the theoretical prediction for small $e$, $\sigma=3e/4$.}
  \label{3}
\end{figure}

In Fig.~\ref{3} we plot the growth rate of the fastest growing mode maximised over $k_{\xi}$ in the range $k_{\xi}\in [0,5]$ as a function of $e$. This range in $k_{\xi}$ was chosen to limit to the computational cost of a wide parameter search, and was found to be sufficient to capture the fastest growing mode in all cases. The numerical results are shown as blue circles and the analytical prediction is shown as a black dashed line with slope $3e/4$.  The small-$e$ analytical prediction for the growth rate at exact resonance correctly captures the instability until $e\gtrsim0.4$, above which the numerically determined growth rate is found to deviate. The peak growth rate is no longer independent of $k_{\xi}$ when $e\not \ll 1$ (see Fig.~\ref{1}).

The laminar flows were found to have quite extreme behaviour for $e\gtrsim 0.4$ or so, with an asymmetric character such that there are strong compressions occurring very close to pericentre (OB14). That these laminar flows differ from simple sinusoidal behaviour for moderately large eccentricities could reduce their ability to excite inertial waves, and might explain why the growth rate is smaller than would be predicted from a simple extrapolation of the small-$e$ behaviour for $e\gtrsim0.4$. Nevertheless, the growth rate in these cases is still large enough for the instability to be dynamically important within a few orbits. For moderate $e$, instability is possible for any $k_{\xi}>0$.

Note that the maximum growth rate for the local instability in a uniformly eccentric isothermal disc is much stronger than the corresponding growth rate obtained by \cite{John2005a}. He found that in a cylindrical disc model (without vertical structure), $\sigma = 3 e/16$ in the limit that $k_{\xi},n\rightarrow \infty$. The difference between these follows from our inclusion of the laminar vertical oscillation of the disc, which provides an additional periodic forcing, and an additional free energy source. The vertical disc oscillations are thus able to amplify the growth rate of the instabilities. We have confirmed that we obtain the analytical prediction of \cite{John2005a} if we artificially neglect the laminar flows by choosing $w=g-1=0$ (see Appendix \ref{Theory}). This highlights the importance of considering the three-dimensional structure of an eccentric disc to correctly capture the instability.

\subsection{Circular reference orbit with nonzero eccentricity gradient}
\label{egradient}
The next case to consider turns out in fact to be the simplest: the instability of a disc that is locally circular but has a nonzero eccentricity gradient. In this case, $w=g-1=0$ for an isothermal disc, so that the vertical laminar flows are no longer present, and there are no corresponding couplings between different $n$ in Eqs.~\ref{EqH1}--\ref{EqH4}. An instability is driven by the periodic variation of the orbital motion of the gas on orbits that neighbour our reference circular orbit. This is analysed in Appendix \ref{Theory}, and is found to have the same character as the instability described in \S \ref{uniforme}.

\begin{figure*}
  \begin{center}
       \subfigure[$|\lambda e^{\prime}|=0.01$]{\includegraphics[trim=6cm 0cm 6cm 0cm, clip=true,width=0.49\textwidth]{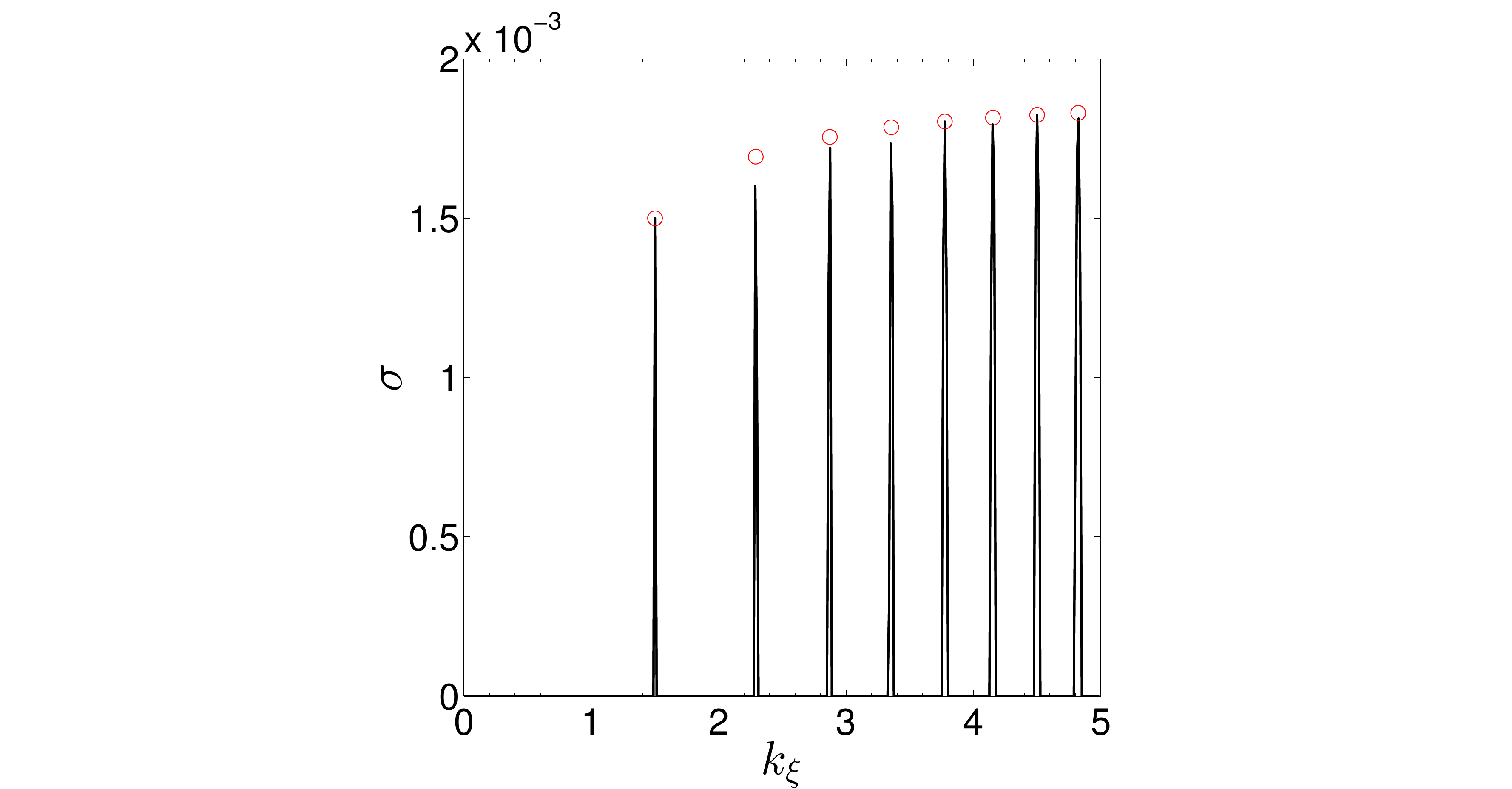} } 
       \subfigure[$|\lambda e^{\prime}|=0.05$]{\includegraphics[trim=6cm 0cm 6cm 0cm, clip=true,width=0.49\textwidth]{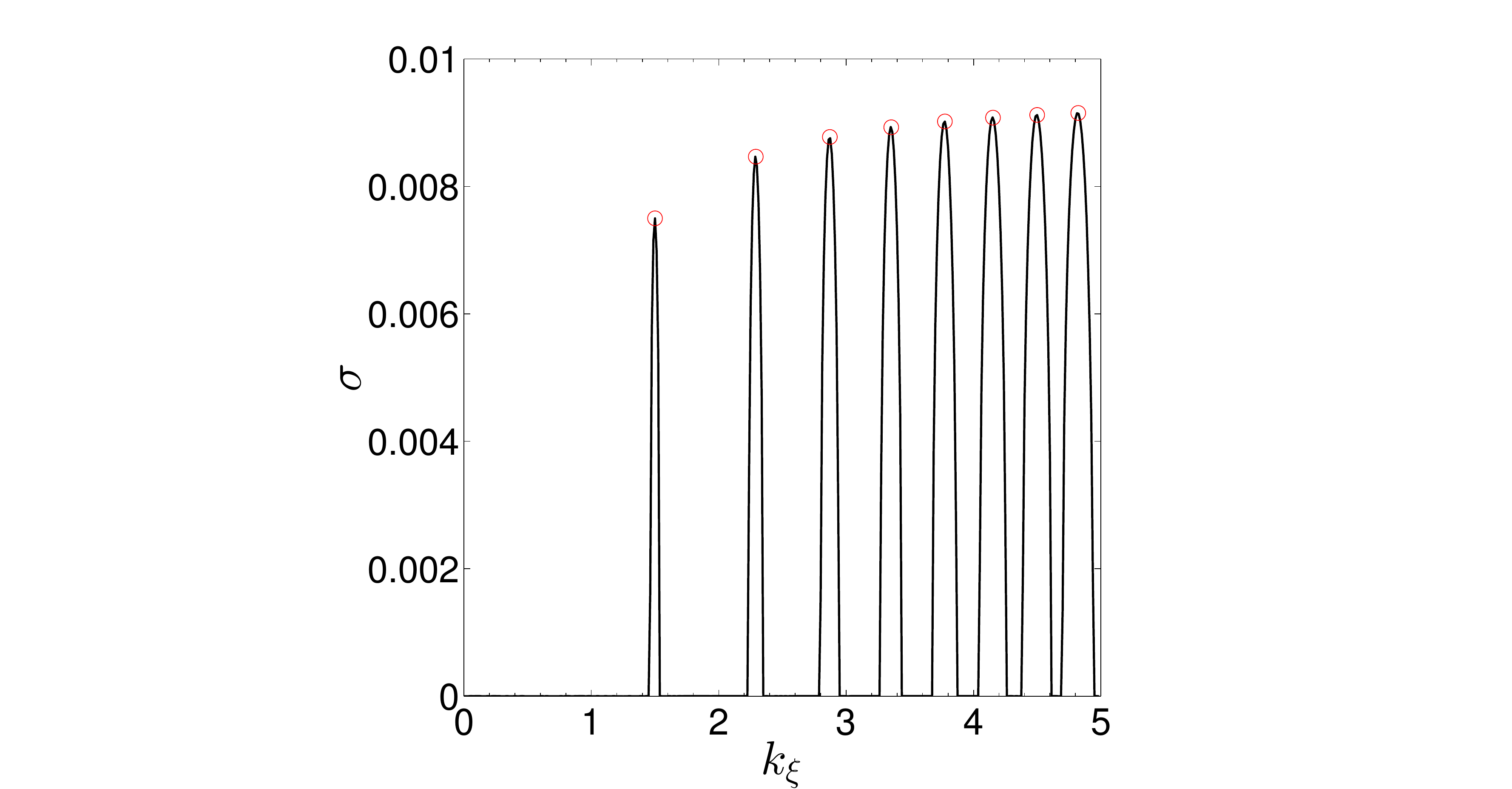} } 
       \subfigure[$|\lambda e^{\prime}|=0.1$]{\includegraphics[trim=6cm 0cm 6cm 0cm, clip=true,width=0.49\textwidth]{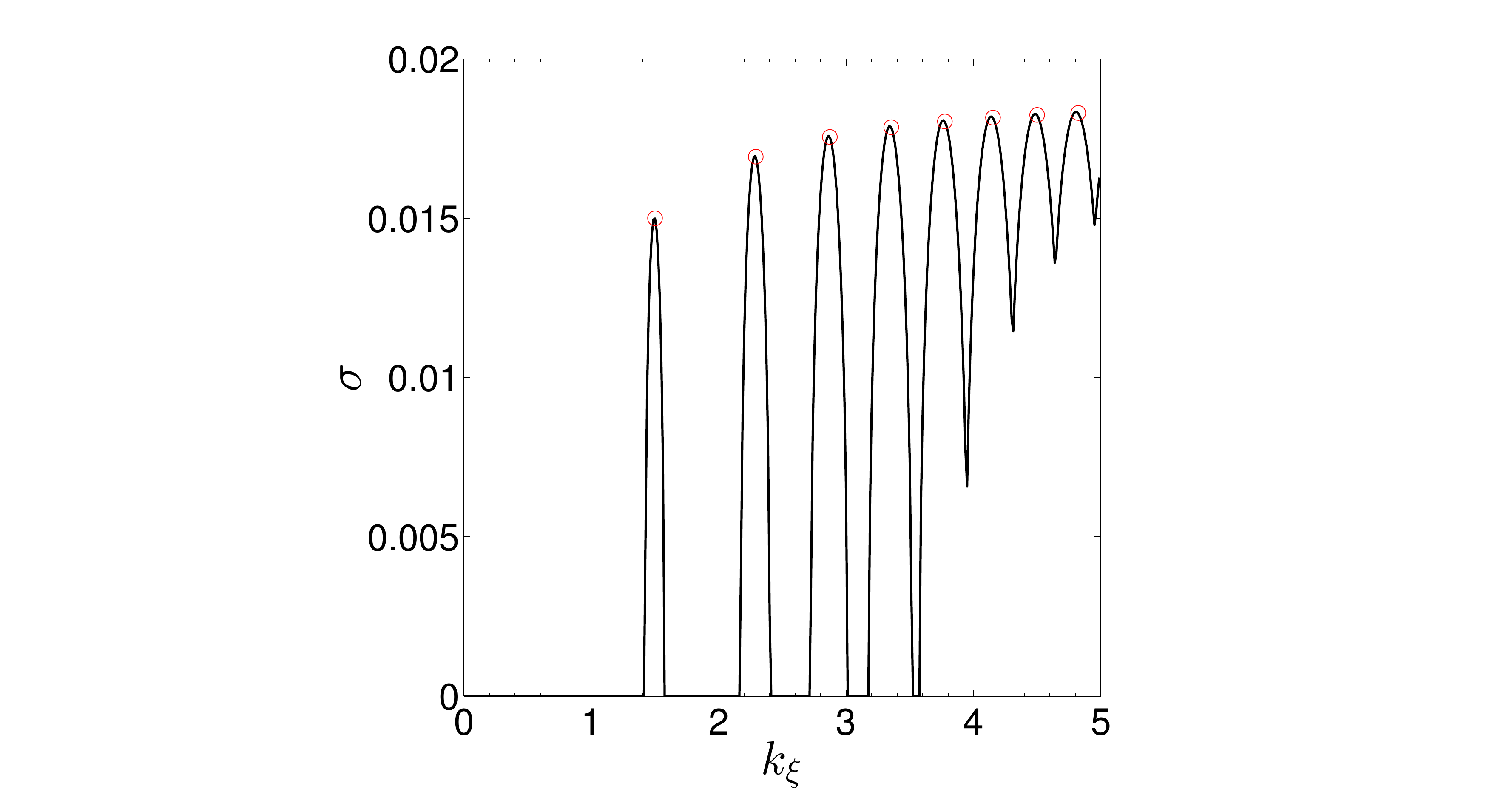} } 
       \subfigure[$|\lambda e^{\prime}|=0.2$]{\includegraphics[trim=6cm 0cm 6cm 0cm, clip=true,width=0.49\textwidth]{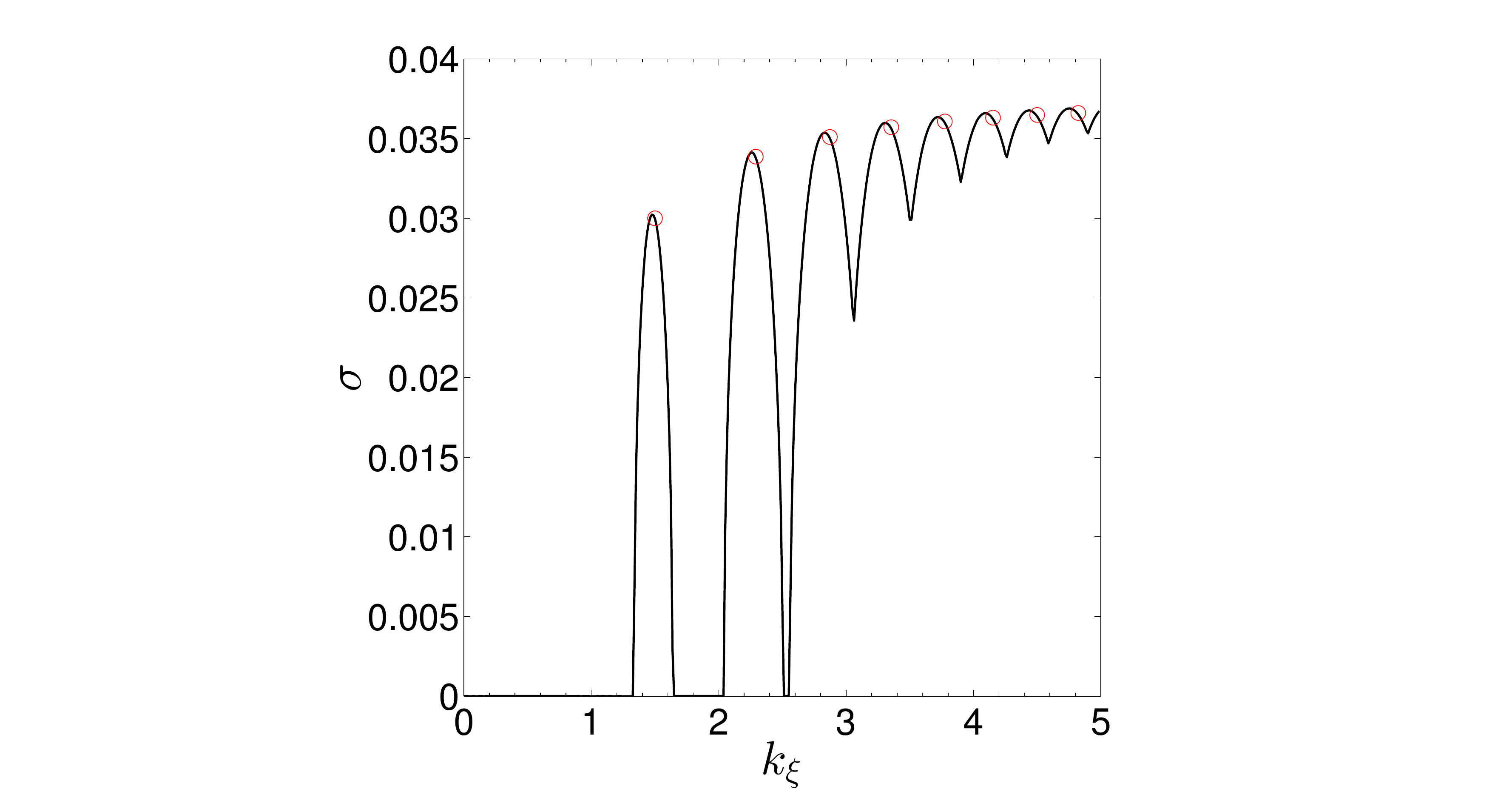} } 
        \subfigure[$|\lambda e^{\prime}|=0.5$]{\includegraphics[trim=6cm 0cm 6cm 0cm, clip=true,width=0.49\textwidth]{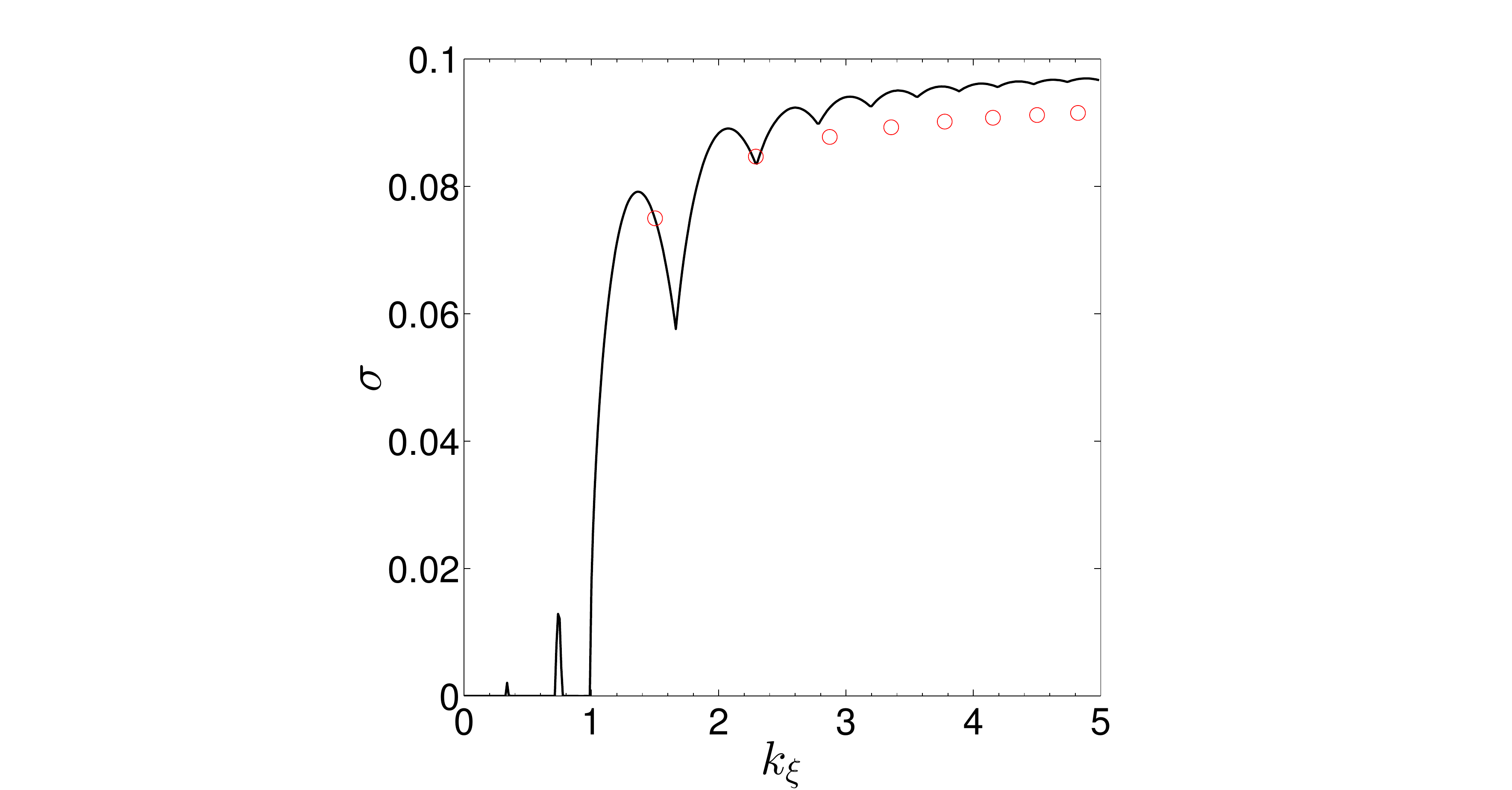} }
         \subfigure[$|\lambda e^{\prime}|=0.99$]{\includegraphics[trim=6cm 0cm 6cm 0cm, clip=true,width=0.49\textwidth]{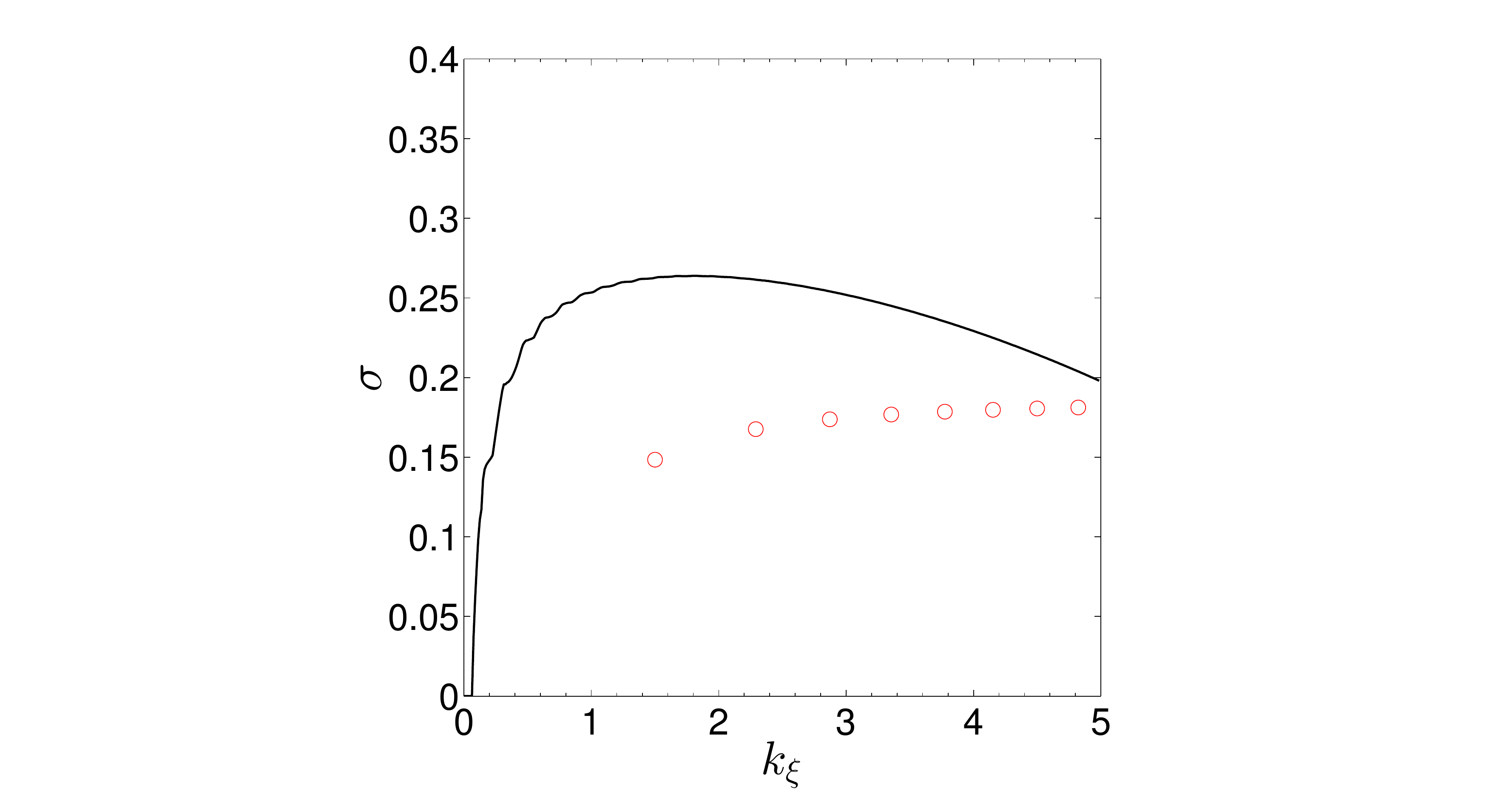} }
    \end{center}
  \caption{Instability in the case of a circular reference orbit with $e=0$ and a nonzero eccentricity gradient. The growth rate of the fastest growing mode (using units such that $\left(GM/\lambda_{0}^3\right)^{\frac{1}{2}}=1$) is plotted as a function of $k_{\xi}$ for various $|\lambda e^{\prime}|$. The red circles show the analytical prediction from Appendix \ref{Theory} at exact resonance, which is valid when $|\lambda e^{\prime}|\ll 1$. The numerical calculations were performed up to $N=12$, which was sufficient to obtain the fastest growing mode in all cases, and involved calculations at $400$ uniformly distributed values of $k_{\xi}\in[0,5]$. Note that $|\lambda e^{\prime}|< 1$ for non-intersecting orbits.}
  \label{4}
\end{figure*}

\begin{figure}
  \begin{center}
  \subfigure{\includegraphics[trim=6cm 0cm 6cm 0cm, clip=true,width=0.49\textwidth]{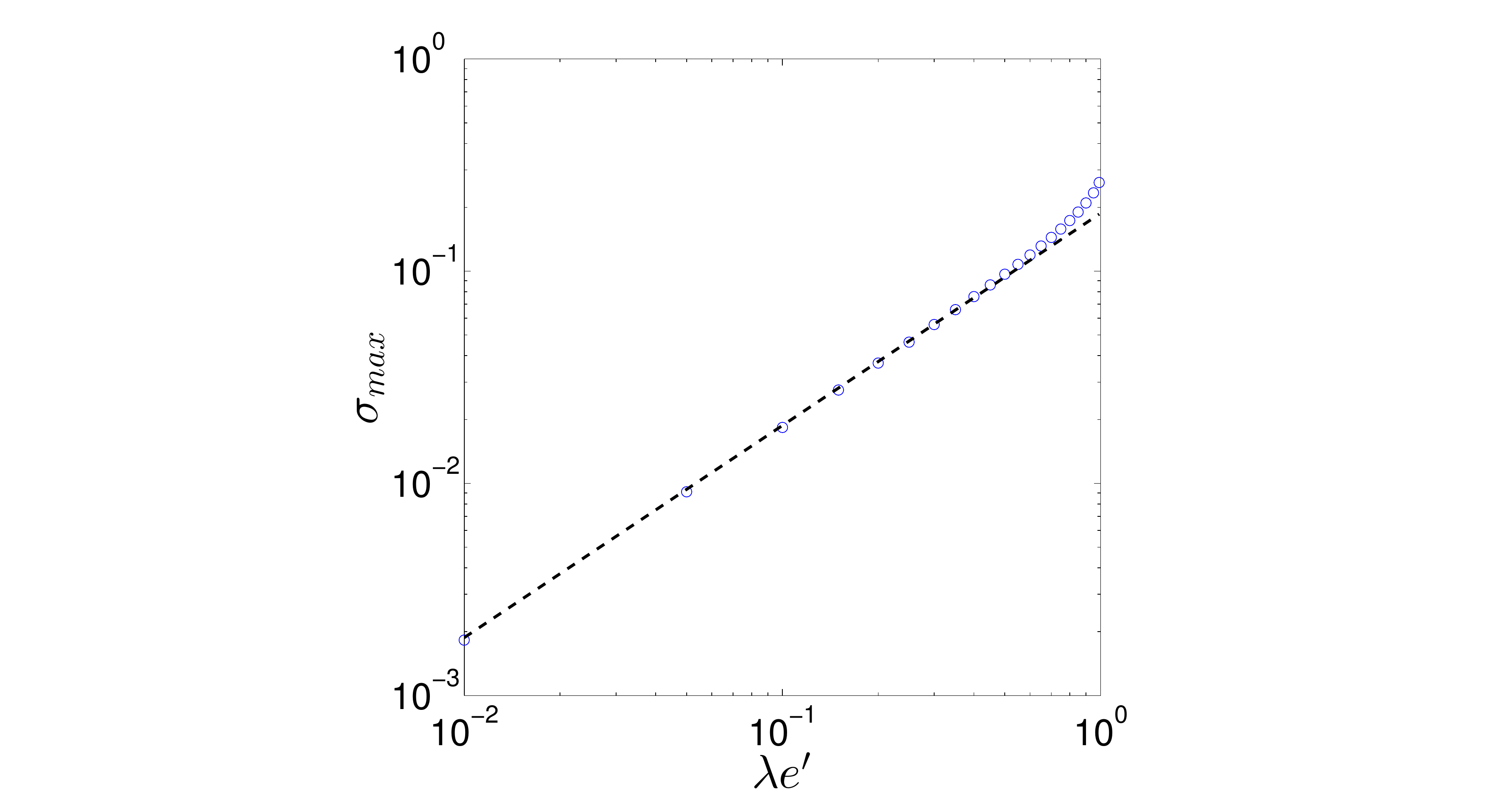} }
    \end{center}
  \caption{Maximum growth rate of the instability (for $k_{\xi}\in [0,5]$) as a function of $|\lambda e^{\prime}|$ when $e=0$ (using units such that $\left(GM/\lambda_{0}^3\right)^{\frac{1}{2}}=1$). The numerical results are plotted as blue circles, and were computed up to $N=12$, which was found to be sufficient to obtain the fastest growing mode in all cases. The black dashed line is the theoretical prediction for small $|\lambda e^{\prime}|$, $\sigma_{max}=3|\lambda e^{\prime}|/16$.}
  \label{5}
\end{figure}

In Fig.~\ref{4} we plot the growth rate of the fastest growing mode from our numerical calculations as a function of the radial wavenumber $k_{\xi}$ for various $|\lambda e^{\prime}|$. We have also plotted our analytical predictions from Appendix \ref{Theory} for the growth rate at exact resonance as red circles, where for small $|\lambda e^{\prime}|$, $\sigma=3 |\lambda e^{\prime}| /16$ for large $k_{\xi},n$. In this case the instability is somewhat weaker than that for a uniformly eccentric disc because the additional energy source provided by the vertical oscillations of the disc is absent. This means that for the smallest $|\lambda e^{\prime}|$ that we have considered, the instability bands are very narrow (in the first panel, exact resonance is not captured by our distribution of points in $k_{\xi}$). However, for large $|\lambda e^{\prime}|$, instability is possible for any $k_{\xi}>0$.
Note that non-intersecting orbits must have $|\lambda e^{\prime}|<1$ (if this is violated, the instability that we have described will no longer be relevant, since the flow will develop shocks).

In Fig.~\ref{5} we plot the growth rate of the fastest growing mode maximised over $k_{\xi}$ in the range $k_{\xi}\in [0,5]$ as a function of $|\lambda e^{\prime}|$. The numerically determined values are shown as blue circles and the black dashed line shows the small-$|\lambda e^{\prime}|$ analytical prediction as $k_{\xi},n\rightarrow \infty$. The analytical prediction works well until $|\lambda e^{\prime}| \sim 1$, near to which the growth rate is slightly amplified over the small-$|\lambda e^{\prime}|$ prediction.

\subsection{General eccentric disc}

\begin{figure*}
  \begin{center}
       \subfigure{\includegraphics[trim=0cm 0cm 0cm 0cm, clip=true,width=0.48\textwidth]{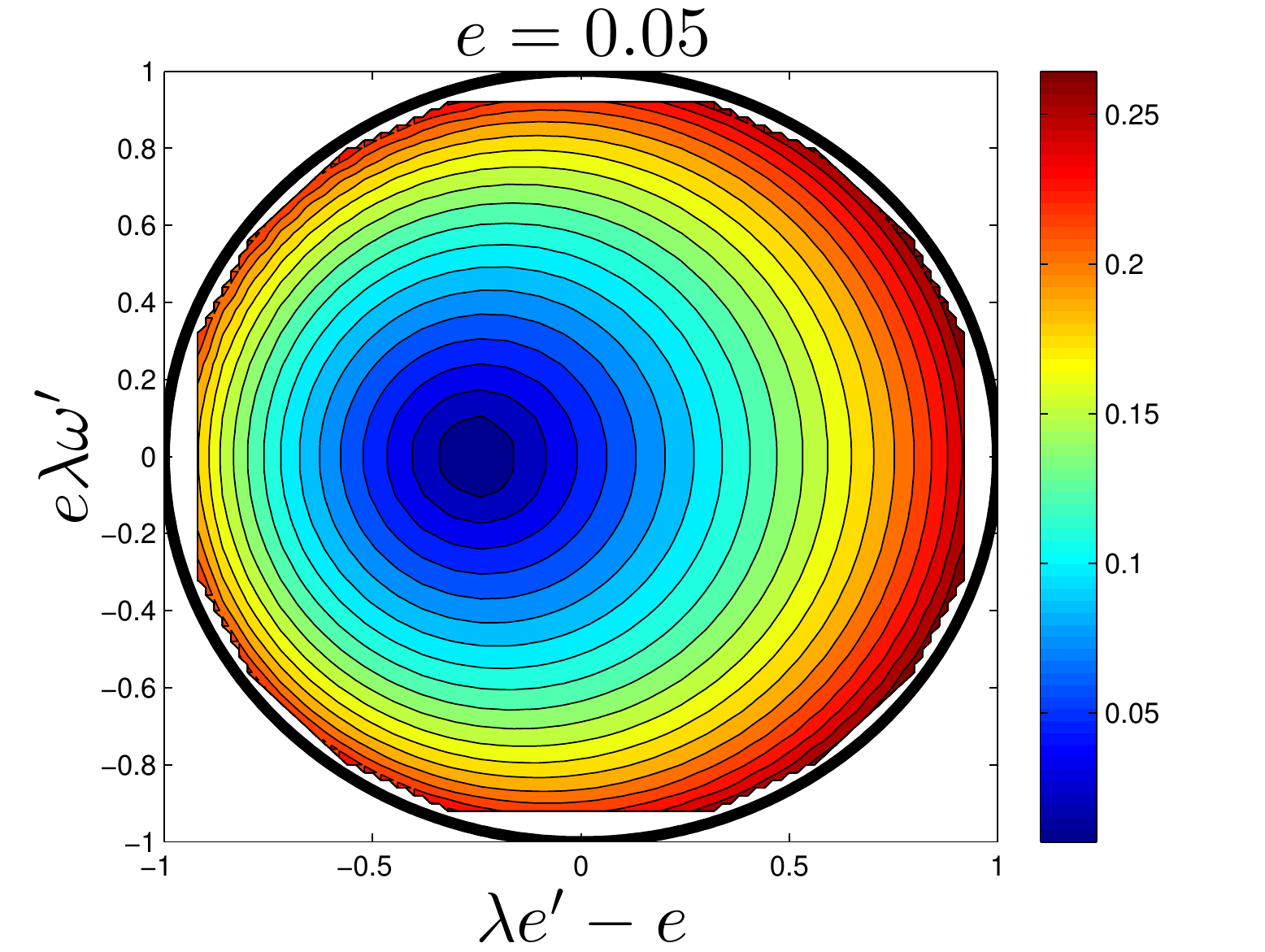} }
        \subfigure{\includegraphics[trim=0cm 0cm 0cm 0cm, clip=true,width=0.48\textwidth]{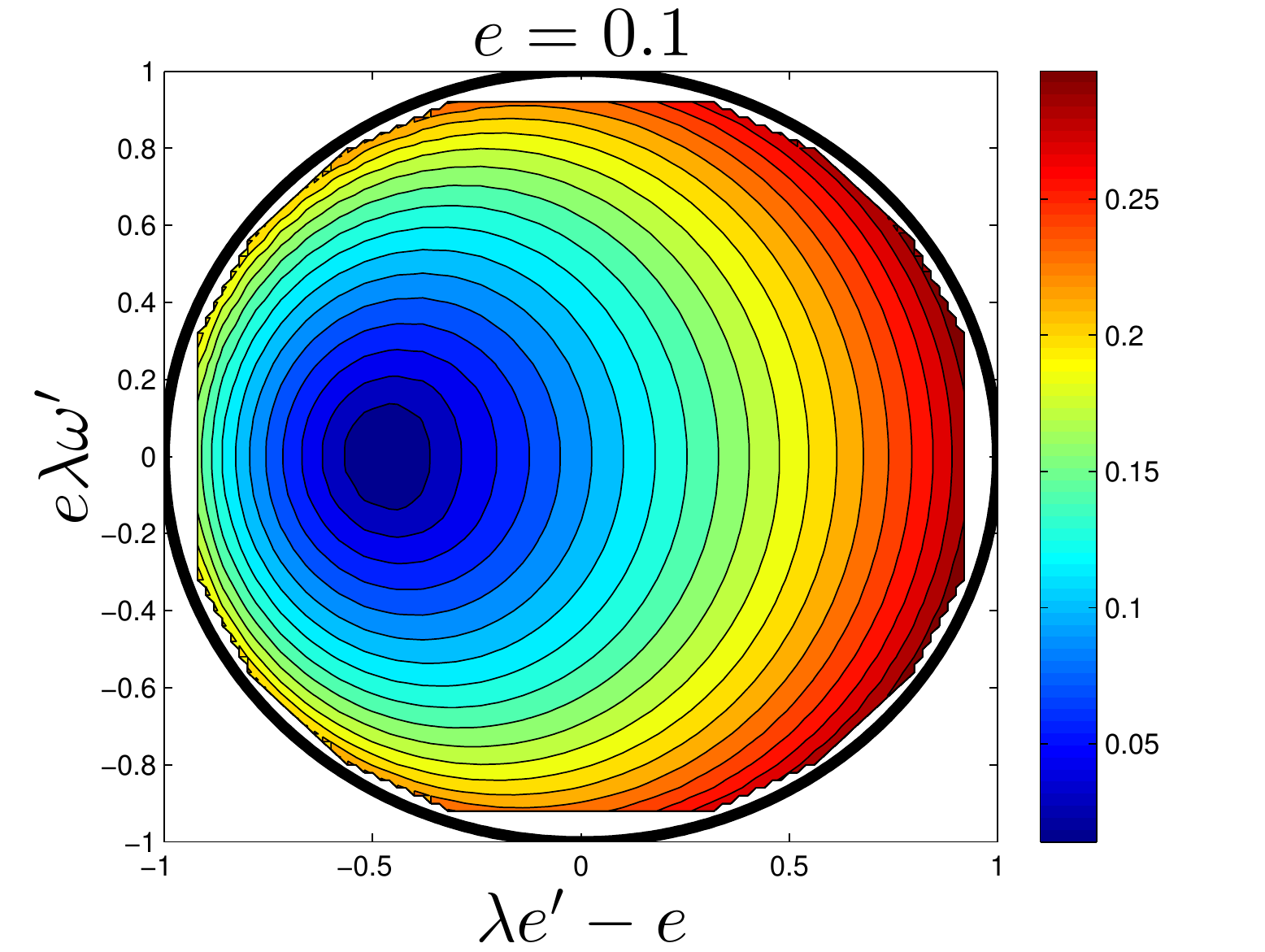} }
        \subfigure{\includegraphics[trim=0cm 0cm 0cm 0cm, clip=true,width=0.48\textwidth]{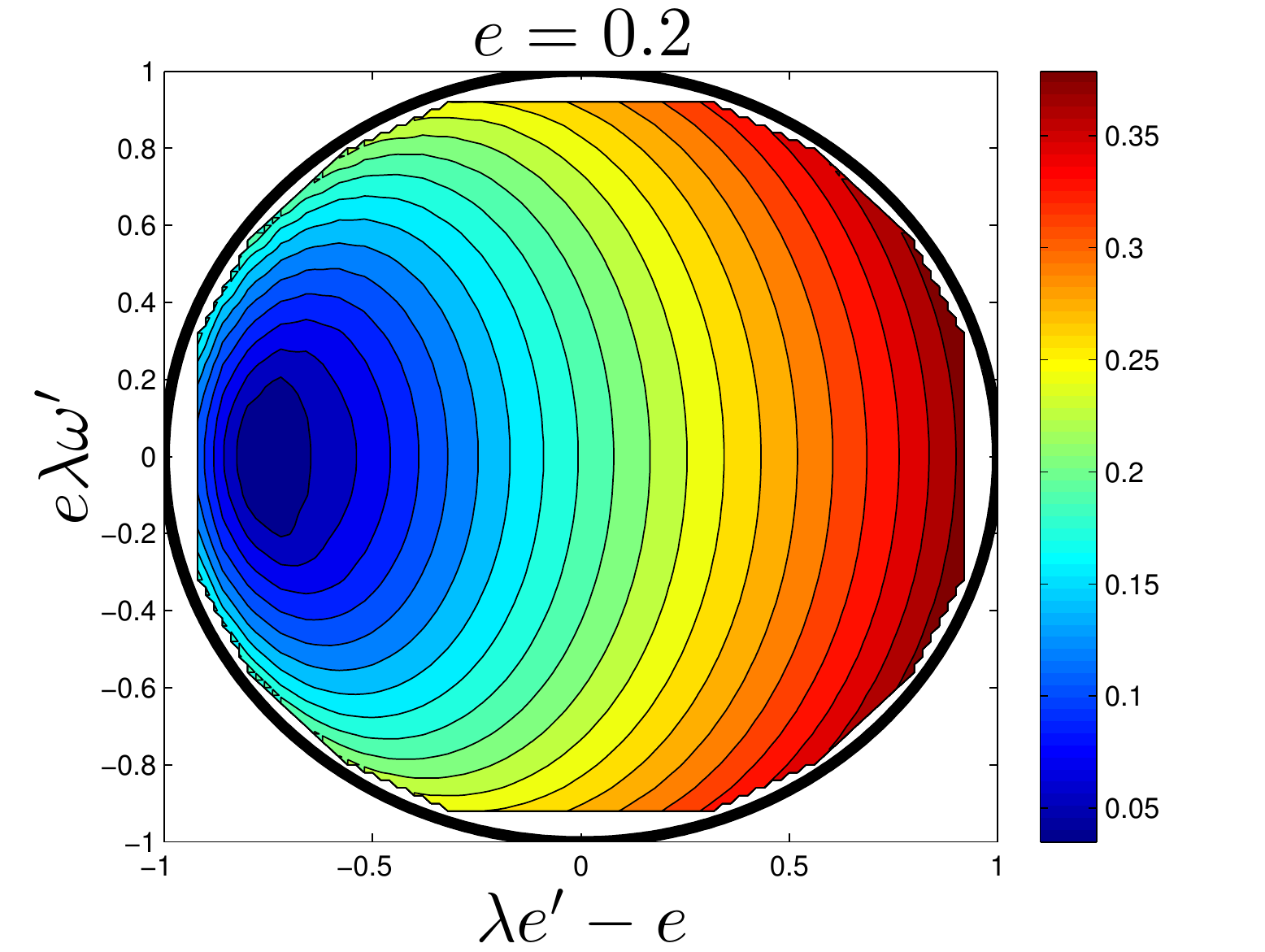} }
         \subfigure{\includegraphics[trim=0cm 0cm 0cm 0cm, clip=true,width=0.48\textwidth]{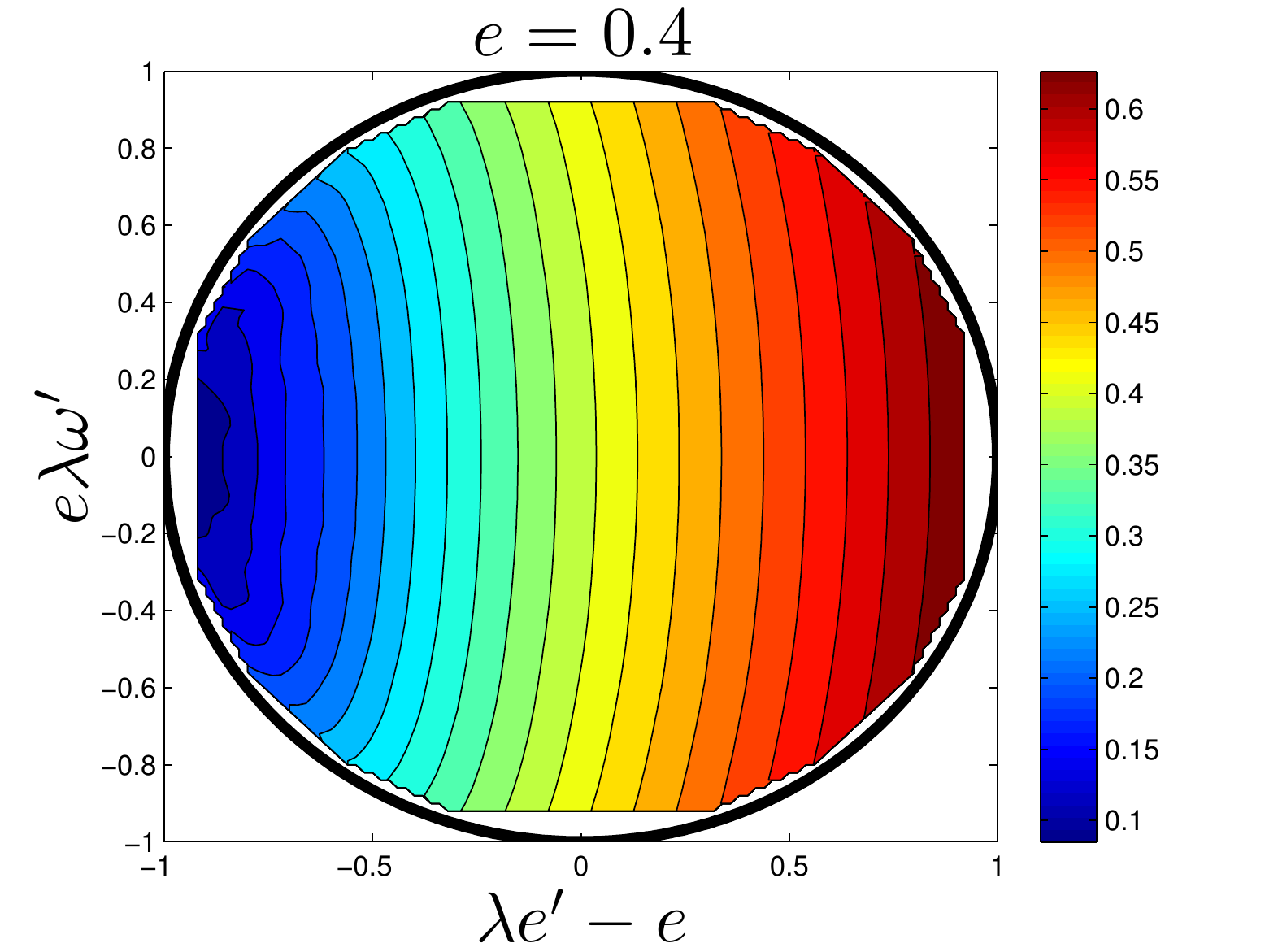} }
           \subfigure{\includegraphics[trim=0cm 0cm 0cm 0cm, clip=true,width=0.48\textwidth]{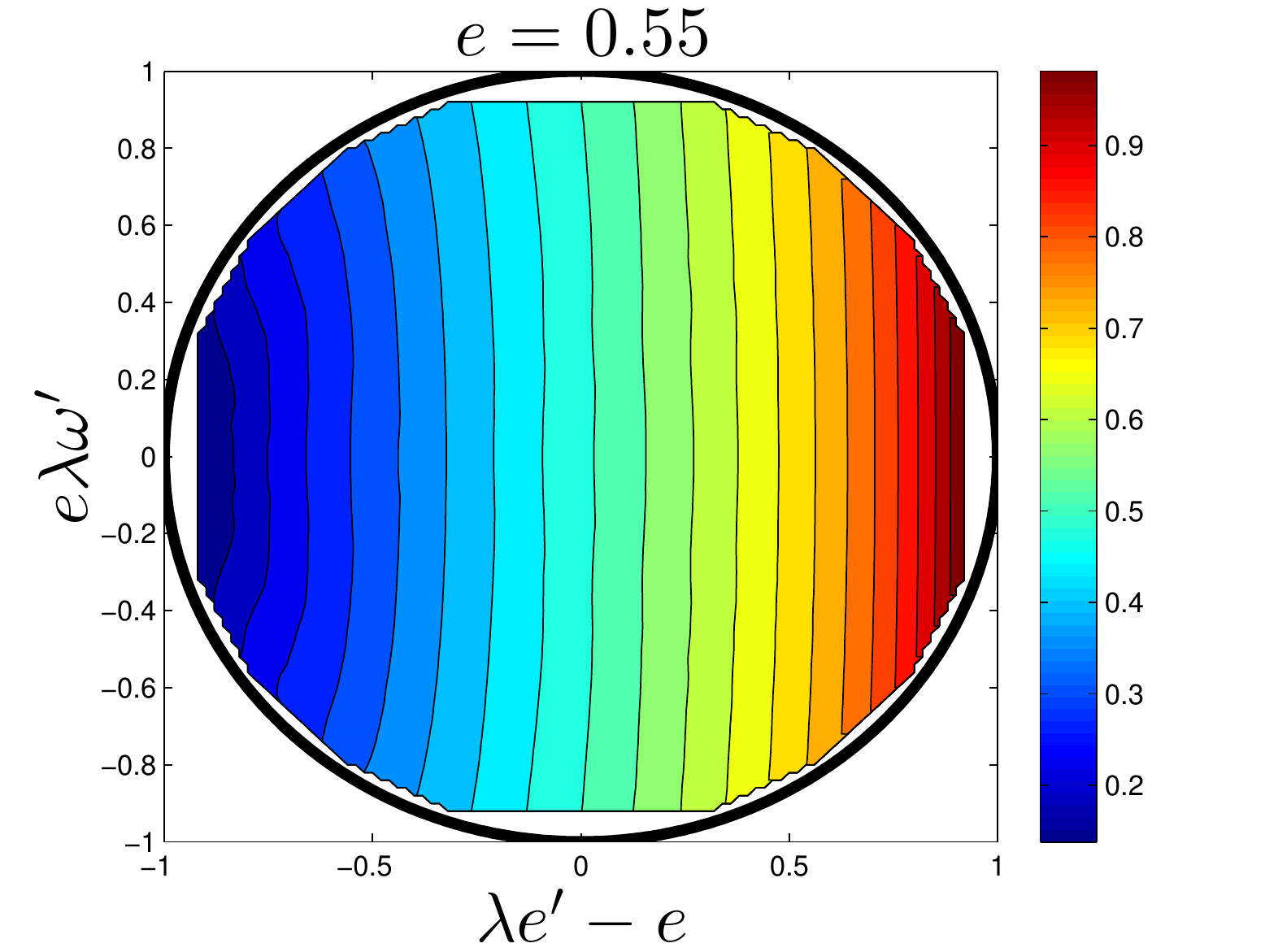} }
       \end{center}
  \caption{Growth rate of the instability (using units such that $\left(GM/\lambda_{0}^3\right)^{\frac{1}{2}}=1$) maximised over $k_{\xi}\in [0,5]$ for a general eccentric disc on the $(\lambda e^{\prime}-e,e\lambda\omega^{\prime})$-plane for several values of $e$. There are 30 points in each coordinate of a Cartesian grid in the range $[-1,1]$, with vertical mode numbers up to $N=9$. The results are then interpolated and smoothed. Orbits outside of the circle of radius 1 are intersecting.}
  \label{6}
\end{figure*}

We will now describe the instability for the more general configuration of an eccentric disc with a nonzero eccentricity gradient. The requirement for orbits to not intersect is
\begin{eqnarray}
\label{intersect}
\left(e-\lambda e^{\prime}\right)^{2}+e^{2}(\lambda\omega^{\prime})^{2}<1.
\end{eqnarray}
If this is violated, we expect shocks to form, and they are likely to dominate the evolution of the disc. We compute the maximum growth rate of the instability and plot its contours on the $(\lambda e^{\prime}-e,e\lambda\omega^{\prime})$-plane for several values of $e$ in Fig.~\ref{6} -- this plane was chosen since the requirement for orbits to be non-intersecting is represented as the region inside the unit circle. This figure illustrates that we have instability over most of the $(e,\lambda e^{\prime},\lambda\omega^{\prime})$ parameter space, and that the growth rate is generally larger when we have larger eccentricities, as well as larger eccentricity gradients. The analytical prediction for this general case for small departures from a circular reference orbit is presented in Appendix \ref{Theory}. We find that the instability of an eccentric disc is stronger for larger\footnote{This may seem somewhat surprising because the orbits intersect when $e-\lambda e^{\prime}$ is sufficiently large. Therefore we might expect the growth rate to increase with $e-\lambda e^{\prime}$. However, inertial waves are not excited as efficiently when $e$ and $\lambda e^{\prime}$ have opposite signs.} $e+\lambda e^{\prime}$. Note, however, that there is a region where the instability is weak, centred on $\lambda \omega^{\prime}=0$. For large $n$ the growth rate is zero when $\lambda e^{\prime}=-4e$ for small $e$ (though the growth rate is not exactly zero for any finite $n$), which is predicted by the analysis in Appendix \ref{Theory} (departures from this prediction in Fig.~\ref{6} are apparent for larger $e$, where this region moves further to the left of the allowed parameter space). For those particular choices of parameters, the coupling between the eccentric disc motion and the inertial waves is weak. However, instability is possible over nearly all of the parameter space for an eccentric disc.

For the case $e=0$, which was described in \S \ref{egradient}, the only relevant parameter describing the orbit is $\lambda e^{\prime}$. In this case, if we were to plot this on Fig.~\ref{6}, the contours would be perfect circles centred on the origin, with the growth rate shown in Fig.~\ref{5}. As $e$ is increased, the growth rate contours begin to differ from circles and when $e\gtrsim 0.4$, they become increasingly independent of $\lambda \omega^{\prime}$ i.e.~they are better approximated by vertical lines. This presumably results from the increasingly non-sinusoidal behaviour of the laminar flow solutions, which become more strongly localised near pericentre for moderately large $e$. Therefore they may not be as efficient at exciting inertial waves.

Fig.~\ref{6} illustrates that instabilities growing on a dynamical timescale are possible for an orbit with any eccentricity considered, as long as the eccentricity gradient is sufficiently large. This also shows that the instability of an eccentric disc is widespread.

\section{Neutrally stratified polytropic discs: dependence on the adiabatic index}

The results presented in \S \ref{Numerical} were obtained by assuming an isothermal relation, which is the most compressible model that we can adopt. Given that the compressibility of the disc plays an important role in driving the laminar vertical oscillations, and that the presence of these oscillations was found to significantly amplify the growth rate of the instability, it is important to determine how these results depend on the adiabatic index. In this section we describe our analytical results for a neutrally stratified polytropic disc which behaves adiabatically with $p\propto \rho^{\gamma}$, where $\gamma=1+1/n_{p}$, and $n_{p}$ is the polytropic index. Realistic discs are expected to have $\gamma\approx 1.4-1.7$. We neglect any possible stable (or unstable) vertical stratification since this requires the eigenfunctions to be localised near to the mid-plane of the disc, and this would complicate the analysis.

In Appendix \ref{WKBTheory} we present the analytical theory describing the locally axisymmetric instability of an eccentric disc using a WKB approximation, in which the radial and vertical wavelengths of the unstable mode are taken to be much smaller than the disc thickness. Looking for such small-scale instabilities allows us to treat the unstable modes locally as plane waves, which avoids the complication that there are no analytically computable eigenmodes for the circular polytropic disc, unlike the isothermal disc considered so far. The WKB theory in Appendix \ref{WKBTheory} extends the calculations of \cite{John2005a} to include the vertical structure of the disc, and to allow for any eccentricity gradient and adiabatic index (for a neutrally stratified disc).

The instability is found to take the same form for any $\gamma$, and involves the excitation of pairs of inertial waves with $\omega=\pm1/2$ that form a standing wave. However, the growth rate of the instability is found to depend on $\gamma$. In particular, the instability for a uniformly eccentric disc has a growth rate
\begin{eqnarray}
\sigma = \frac{3}{16}\left(1+\frac{3}{\gamma}\right)e,
\end{eqnarray}
which is strongest for an isothermal disc ($\gamma=1$ leads to $3e/4$) and reduces to $3e/16$ for an incompressible disc ($\gamma\rightarrow \infty$). An incompressible disc does not exhibit vertical laminar oscillations in this case, so the instability is driven purely by the periodic variation of the eccentric orbital motion around an orbit. Hence the agreement with the growth rate obtained by \cite{John2005a}. This demonstrates once again that the vertical laminar flows play an important role in driving instabilities in an eccentric disc. Note that the growth rate is significantly amplified over the incompressible limit for any realistic $\gamma$. For example, $\sigma=21\, e/40\approx0.525\, e$ when $\gamma=\frac{5}{3}$, which is appropriate for a disc consisting of ionised hydrogen.

Similarly, if we consider the instability of a circular reference orbit with a nonzero eccentricity gradient, we obtain
\begin{eqnarray}
\sigma =\frac{3}{16\gamma}|\lambda e^{\prime}|,
\end{eqnarray}
which is strongest for an isothermal disc ($3|\lambda e^{\prime}|/16$), and vanishes entirely in the incompressible limit.

The more general configuration of an eccentric disc with a nonzero eccentricity gradient is presented in Appendix \ref{WKBTheory}.

\section{Discussion}

The local instability that we have analysed in this paper occurs whenever an astrophysical disc becomes eccentric. Assuming a simple $\alpha$-prescription for the turbulence in a Keplerian disc, we can simply estimate to what degree a departure from circularity is required for the instability to grow in the presence of viscosity. The viscous damping rate of a mode with radial wavenumber $k_{\xi}$ and vertical mode number $n$ is approximately\footnote{The viscous linearised equations for the instabilities in a warped disc were studied more carefully in \citealt{OL2013b}.} $\alpha(k_{\xi}^2+n)$, where $\alpha$ is the viscosity coefficient. For a uniformly eccentric disc, we require $e\gtrsim 0.04 (\alpha/10^{-2})$ for the largest wavelength ($n=1,k_{\xi}=3/2$) instability to occur. Similarly, for a circular reference orbit with a nonzero eccentricity gradient, we require $|\lambda e^{\prime}| \gtrsim 0.17(\alpha/10^{-2})$ for instability. Note, however, that instability is strongest for a combination of eccentricity and eccentricity gradient, therefore instability may be possible even if these criteria are not satisfied. In addition, it is unclear whether the interaction of this instability with the turbulence that drives accretion in the disc can be modelled with a simple $\alpha$-viscosity prescription. Nevertheless, we conclude that instability is possible in a sufficiently eccentric disc for typical values of $\alpha$ thought to be relevant for circumstellar discs.

Another aspect is that the local instability involves the coupling of pairs of inertial waves that propagate radially in opposite directions in the disc to form a local standing wave. If the disc eccentricity or its gradient is localised to some region of the disc of radial extent $D$ (the ``interaction region"), the instability will only cause disturbances to reach large amplitudes if the waves spend enough time in that region to be sufficiently amplified. We expect the instability to be important if the growth time is shorter than the wave crossing time over this region (unless the waves can reflect from radial boundaries and re-enter the interaction region), which is approximately $\sigma^{-1} \lesssim D/c_{g}$, where the group velocity of an inertial wave is $c_{g}\sim \omega/k_{\xi}$. This suggests that $k_{\xi}\gtrsim O((e D)^{-1})$ is required for the instability to amplify disturbances to large amplitudes, so that a small interaction region or a weak eccentricity will preferentially excite small-scale disturbances, which may be more difficult to observe in simulations, or be more easily damped by viscosity (this simplistic argument neglects the presence of an eccentricity gradient).

Previous grid-based hydrodynamical simulations of disc-companion tidal interactions have observed the generation of local eccentricity and eccentricity gradients in the disc (e.g.~\citealt{PapNM2001,KleyDirksen2006,DAngelo2006,KleyPO2008,Marzari2012}). However, the instabilities that we have described in this paper have never been observed previously\footnote{Except by \cite{John2005b}, who performed a set of global simulations specifically designed to study them for the specific case of a cylindrical disc (lacking vertical structure).}. This is partly because most of the existing simulations are two-dimensional, therefore they would be unable to capture the instability. In addition, these simulations would also incorrectly neglect the vertical laminar oscillations of an eccentric disc. The limited three-dimensional simulations that have been performed thus far (e.g.~\citealt{Bitsch2013}) necessarily have limited spatial resolution (or too large a physical or numerical viscosity), so they may have been unable to capture the instability\footnote{The simulations in \cite{Bitsch2013} of an isothermal vertically structured disc have $\alpha=0.005$. The eccentricity of the disc attains values of approximately $e\sim 0.3$ and $|\lambda e^{\prime}|\sim 0.5$, so the largest wavelength instability can in principle be excited. However, the radial extent of the region of moderate disc eccentricity is $\lesssim 1$, so the most strongly excited waves will have smaller radial wavelengths, with $k_{\xi}\gtrsim 6, n\gtrsim 10$, which will most likely be damped by the viscosity adopted. Hence it is not surprising that this instability is not observed in their simulations.}. In addition, the instability may be too weak to be observed during the limited duration of some simulations, particularly when the disc eccentricity or eccentricity gradients do not attain large values.

In a similar way, global SPH simulations of eccentric discs in superhump binaries are either two-dimensional \citep{Whitehurst1988,Lubow1991b}, or they do not have sufficient spatial resolution to be able to capture these instabilities \citep{Smith2007}. Previous simulations have captured the exponential growth of eccentricity due to the tidal instability of \cite{Whitehurst1988} and \cite{Lubow1991a}, but they do not identify a mechanism of saturation for the instability. The instabilities that we have analysed in this paper may provide such a mechanism, because their growth rates are an increasing function of the eccentricity.

These instabilities may play a role in modifying the eccentricities of planets undergoing tidal interactions with the protoplanetary disc. Secular interactions between the planets and the disc will redistribute the angular momentum deficit between the various components, leading to oscillations in the disc and planet eccentricities (e.g.~\citealt{PapNM2001,Pap2002,GoldreichSari2003}). When parts of the disc become eccentric, we speculate that the instabilities analysed here would lead to a decay of the disc eccentricity and eccentricity gradient (this behaviour was observed by \citealt{John2005b}). These instabilities would therefore reduce the angular momentum deficit of the system, and provide a mechanism for damping the eccentricities of the planets due to their secular coupling with the disc. To determine the efficiency of this damping process, nonlinear calculations are required, which we defer to future work. 

\section{Conclusions}
In this paper we have studied the hydrodynamic stability of eccentric Keplerian discs. We have utilised a local model similar to the conventional shearing box, which we derived in a companion paper (OB14), to compute the vertical oscillations of the eccentric disc, and to analyse their resulting instabilities using a numerical Floquet method. We have obtained detailed analytical understanding of the instability for isothermal discs (Appendices \ref{Theory} and \ref{Energy}), as well as for polytropic discs with any polytropic index using a WKB approximation (Appendix \ref{WKBTheory}), for any weak local eccentricity and eccentricity gradient. This work considerably extends the pioneering calculations of \cite{John2005a}, who was the first to identify these instabilities for the specific case of a uniformly eccentric cylindrical disc.

We have highlighted the importance of considering three-dimensional effects and the disc vertical structure in order to understand the evolution of eccentric discs. This arises because vertical oscillations of the disc are driven by the periodic variation in the vertical gravitational acceleration around an eccentric orbit (identified by \citealt{Ogilvie2001}). These oscillations provide an additional free energy source, and an additional periodic driving, of small-scale inertial waves. Instabilities with dynamically relevant growth rates are found in a disc with a sufficiently large local eccentricity or eccentricity gradient. Two-dimensional calculations would be unable to capture either of these effects, and are unlikely to correctly describe the evolution of eccentric discs.

Disc-planet interactions can generate local disc eccentricity and eccentricity gradients \citep{PapNM2001,KleyDirksen2006,DAngelo2006,Bitsch2013}. The secular interaction of planets with eccentric disc modes leads to an exchange of eccentricity between the disc and any planets orbiting within. The instabilities that we have analysed could damp the eccentric modes in the disc. This could provide a mechanism to damp the eccentricities of planets that interact with their discs.
 
The instability that we have studied in this paper is related to the elliptical instability in fluid dynamics (e.g.~\citealt{Kerswell2002}), which is thought to be excited in tidally deformed discs in binary systems \citep{Goodman1993,LubowPringleKerswell1993}, as well as the fluid interiors of stars and giant planets \citep{BL2013a,BL2013b}. The nonlinear evolution of these instabilities in a local model of a tidally deformed disc was studied by \cite{RyuGoodman1994}, who found that they resulted in sustained turbulence (or wave activity) and a tidal torque, together with some weak angular momentum transport. \cite{John2005b} studied the evolution of the instabilities presented in \cite{John2005a} in a global cylindrical Keplerian disc with a free eccentricity. He found that these instabilities lead to a gradual decay of the disc eccentricity. In order to determine the astrophysical implications of these instabilities, it is essential to perform three-dimensional nonlinear numerical simulations of eccentric discs, either using the local model derived in OB14, or in global simulations of discs with vertical structure. We defer such calculations to future work.

\section{Acknowledgments}
This research was supported by STFC through grants ST/J001570/1 and ST/L000636/1.

\appendix

\onecolumn
\section{Geometrical coefficients}
\label{Coefficients}
It is much simpler to use the true anomaly $\theta$ rather than time for studying the local dynamics of an eccentric disc. This is defined by $\theta(\lambda)=\phi-\omega(\lambda)$, where $\omega(\lambda)$ is the longitude of pericentre and $\phi$ is the azimuthal angle. Here we list the coefficients that are relevant for understanding the local instabilities of an eccentric disc. Below, primes denote differentiation with respect to $\lambda$, while $c$ and $s$ denote $\cos\theta$ and $\sin\theta$, respectively. $J$ is the Jacobian of the orbital coordinates, $g_{ij}$ are the components of the metric tensor (and its inverse $g^{ij}$), $\Gamma^{i}_{jk}$ are the components of the Levi-Civita connection, and $\Delta=(1/J)\partial_{\phi}(J\Omega)$ is the orbital velocity divergence (see OB14 for further details).
\begin{eqnarray}
R &=& \lambda (1+e c)^{-1} \\
\Omega &=& \sqrt{\frac{GM}{\lambda^{3}\Omega_{0}^{2}}}(1+e c)^{2} \\
\Phi_{2} &=& (1+e c)^3 \\
J &=&  \lambda \frac{(1-\lambda e^{\prime} c+e c-e\lambda \omega^{\prime} s)}{(1+e c)^3} \\
g^{\lambda\lambda} &=& \frac{(1+e c)^2(1+2 c e +e^2)}{(1-c \lambda e^{\prime}+e c -es\lambda \omega^{\prime})^{2}} \\
\lambda g^{\lambda\phi} &=& \frac{-e s (1+c e)^2}{(1-c \lambda e^{\prime}+e c -es \lambda \omega^{\prime})} \\
\lambda^{2}g^{\phi\phi}&=&(1+e c)^2 \\
g_{\lambda\lambda} &=& \frac{(1-c \lambda e^{\prime}+e c -es\lambda \omega^{\prime})^{2}}{(1+e c)^4} \\
\lambda^{-1} g_{\lambda\phi} &=& e s \frac{(1-c \lambda e^{\prime}+e c -es\lambda \omega^{\prime})}{(1+e c)^4} \\
\lambda^{-2}g_{\phi\phi}&=& \frac{1+2 e c + e^2}{(1+e c )^4} \\
\Gamma^{\lambda}_{\lambda\phi}&=& \frac{(s \lambda e^{\prime} -e(c+e)\lambda \omega^{\prime})}{(1+e c)(1-c \lambda e^{\prime}+e  c- e s \lambda \omega^{\prime})} \\
\lambda^{-1}\Gamma^{\lambda}_{\phi\phi}&=& \frac{-1}{(1-c \lambda e^{\prime}+e  c- e s \lambda \omega^{\prime})} \\
\lambda\Gamma^{\phi}_{\lambda\phi}&=& \frac{(1-c \lambda e^{\prime}+e  c- e s \lambda \omega^{\prime})}{(1+e c)} \\
\Gamma^{\phi}_{\phi\phi}&=& \frac{2 e s}{(1+ e c)} \\
\lambda \partial_{\lambda}\Omega &=& -\frac{3}{2}(1+ec)^{2}+2(1+e c)(c\lambda e^{\prime}+es\lambda \omega^{\prime}) \\
\partial_{\phi}\Omega &=& -2 e s(1+ec) \\
\Delta &=& \frac{(1+e c)(s \lambda e^{\prime}-e(c+e)\lambda \omega^{\prime})}{(1-c \lambda e^{\prime}+e  c- e s \lambda \omega^{\prime})}
\end{eqnarray}

\section{Theory of local parametric instability in an isothermal eccentric disc}
\label{Theory}
In this section we present the theory that explains the parametric instability observed in \S \ref{Numerical} for an isothermal eccentric disc. An instability is possible because the eccentric orbital motion of the gas, together with the periodic vertical oscillations of the disc, couple the waves that exist in an unperturbed circular disc. The approach followed here is similar to the analysis in \cite{OL2013b}, except that the instability (in its simplest form) does not couple modes with different $n$. 

We consider a slightly eccentric disc with a small nonzero eccentricity and eccentricity gradient. We define a small parameter $\epsilon$ such that $e$, $|\lambda e^{\prime}|$  and $e|\lambda \omega^{\prime}|$ are each $O(\epsilon)$. This allows all three parameters to play a role in the instability when $\epsilon\ll1$, and gives the most general expression for the instability growth rate as a function of $(e,\lambda e^{\prime},e\lambda\omega^{\prime})$. We also neglect viscosity and study the instability at exact parametric resonance -- it is straightforward to generalise this calculation to include a slight detuning or damping of the resonance (e.g.~\citealt{OL2013b}).

The laminar flows, which are the solutions of Eqs.~\ref{laminar}--\ref{laminar1}, have the expansion
\begin{eqnarray}
w&=& 3 e s +O(\epsilon^{2}), \\
g&=&1+ 6 e c +O(\epsilon^{2}), \\
f &=& 3 e c + \lambda e^{\prime} c + e\lambda \omega^{\prime} s +O(\epsilon^2).
\end{eqnarray}
Note that $w$ and $g-1$ do not depend on the eccentricity gradient -- this is no longer the case for a polytropic disc with $\gamma \ne 1$ (see Appendix \ref{WKBTheory}).

We employ a multiple-time-scale expansion of the fluid variables such that
\begin{eqnarray}
u^{\xi}_{n}=u^{\xi,0}_{n}(\theta_{0},\theta_{1},\dots)+\epsilon u_{n}^{\xi,1}(\theta_{0},\theta_{1},\dots)+O(\epsilon^{2}),
\end{eqnarray}
and so on for other variables. We define $\theta_{0}=\theta$ and $\theta_{1}=\epsilon\theta$, so that $d_{\theta}=\partial_{0}+\epsilon \partial_{1}+\dots$, etc. The orbital motion varies periodically with $\theta_{0}$, and this drives parametric instabilities that grow on the slow timescale described by $\theta_{1}$. We also define $\boldsymbol{U}_{n}=\left[u^{\xi}_{n},u^{\eta}_{n},u^{\zeta}_{n},h_{n}\right]^{T}$.

Based on the properties of the fastest growing modes observed in our numerical calculations in \S \ref{Numerical}, we study the exact parametric instability of a pair of inertial waves with $\omega=\pm\frac{1}{2}$ and the same $n$, so that $k_{\xi}=\frac{1}{2}\sqrt{3(4n-1)}$. At leading order ($O(\epsilon^{0})$), this pair of linear waves can be written
\begin{eqnarray}
\label{unstablemode}
\boldsymbol{U}_{n}^{0}=A^{+}_{n}(\theta_{1})\hat{\boldsymbol{U}}^{+0}_{n}+A^{-}_{n}(\theta_{1})\hat{\boldsymbol{U}}^{-0}_{n},
\end{eqnarray}
where
\begin{eqnarray}
\hat{\boldsymbol{U}}^{\pm0}_{n} = \left[ \begin{array}{c}
\pm \mathrm{i}\omega(\omega^{2}-n) \\
\frac{1}{2}(\omega^{2}-n) \\
\pm n k_{\xi}\omega \\
\mathrm{i} k_{\xi} \omega^{2}
 \end{array} \right] \mathrm{e}^{\mp \mathrm{i}\omega \theta_{0}}
\end{eqnarray}
are both eigenvectors of the unperturbed circular disc.
 The leading-order equations are
\begin{eqnarray}
L_{n}\boldsymbol{U}^{0}_{n}=\boldsymbol{0},
\end{eqnarray}
where
\begin{eqnarray}
L_{n} = \left( \begin{array}{cccc}
\partial_{0} & -2 & 0 & \mathrm{i}k_{\xi} \\
\frac{1}{2} & \partial_{0} & 0  & 0\\
0 & 0 & \partial_{0} & n \\
\mathrm{i}k_{\xi} & 0 & -1 & \partial_{0} \end{array}
\right),
\end{eqnarray}
since our chosen solution is a linear superposition of eigenvectors.

To first order ($O(\epsilon^1)$), we obtain the following system of ODEs for each $n$:
\begin{eqnarray}
\label{1storder}
L_{n}\boldsymbol{U}^{1}_{n}=\boldsymbol{F}^{1}_{n} +\boldsymbol{G}^{1}_{n+2},
\end{eqnarray}
where the effective forcing vectors $\boldsymbol{F}^{1}_{n}$ and $\boldsymbol{G}^{1}_{n+2}$ can be obtained from the expansions of Eq.~\ref{EqH1}--\ref{EqH4} to $O(\epsilon^{1})$. Note that $\boldsymbol{G}^{1}_{n+2}$ couples mode $n$ with mode $m=n+2$. However, there are no additional couplings to modes with $m<n$, so these are ``one-way" couplings, and the modes with $m<n$ will be slaved to the mode with the maximum $n$. The growth rate of the instability is therefore fully determined by considering only the largest $n$. We may therefore neglect $\boldsymbol{G}^{1}_{n+2}$ to analyse the growth rate of the instability (note also that this term is exactly zero if $e=0$).
 
For a general forcing vector with $\boldsymbol{F}_{n}^{1}=\left[a_{n},b_{n},c_{n},d_{n}\right]^{T}$, the necessary solvability condition for the system of equations at this order is
\begin{eqnarray}
- k_{\xi}\omega \left( -\mathrm{i}\omega a_{n}+2b_{n}\right)+\left(-\omega^{2}+1\right)\left(c_{n}- \mathrm{i}\omega d_{n}\right)=0.
\end{eqnarray}
This condition is required to eliminate the secular terms in Eq.~\ref{1storder}, and leads to a pair of amplitude equations relating $A_{n}^{\pm}$ and their derivatives with respect to $\theta_{1}$:
\begin{eqnarray}
\partial_{1}A_{n}^{\pm}=\frac{\pm 3\mathrm{i}}{4(16n-1)}\left[(4n-1)(e+\lambda e^{\prime} \pm \mathrm{i} e\lambda \omega^{\prime})+12 ne \right]A_{n}^{\mp}.
\end{eqnarray}
The growth rate of the instability at exact resonance is therefore
\begin{eqnarray}
\label{growthrate}
\sigma &=& \frac{3}{4}\frac{1}{16n-1}\sqrt{e^2 (16 n-1)^2 + (4 n-1)^2(\lambda e^{\prime})^2+ 2(4 n-1)(16 n-1) e \lambda e^{\prime}+(4n -1)^2(e\lambda \omega^{\prime})^2} \\
&=& \frac{3}{4}\frac{1}{16n-1}| \left(16n-1\right)E+\left(4n-1\right)\lambda E^{\prime}|\\
&\rightarrow& \frac{3}{16}\sqrt{16 e^2+ (\lambda e^{\prime})^2+8e \lambda e^{\prime}+ (e\lambda \omega^{\prime})^2}, \;\;\;\; \mathrm{as} \;\;\;\; n\rightarrow \infty.
\end{eqnarray}
This prediction is in excellent agreement with the numerically computed growth rates presented in \S \ref{Numerical} when $\epsilon \ll1$. This provides a posteriori justification that couplings between different $n$ are not required to explain the instability in \S \ref{Numerical}. We have plotted this analytical prediction as red circles in Figs.~\ref{1} and \ref{4}. Note that this instability is weak for some combination of the orbital parameters. In particular, for large $n$ when $\lambda \omega^{\prime}=0$, the instability vanishes when $\lambda e^{\prime}=-4e$ (and is non-vanishing but weak for finite $n$). This corresponds with the region of weak instability present in Fig.~\ref{6} (at least when $\epsilon\ll 1$), at which the nonlinear coupling is found to be weak.

Several special cases of this instability can be considered.

\subsection{Uniformly eccentric disc}
For a uniform eccentric disc with $\lambda e^{\prime}=\lambda \omega^{\prime}=0$, we find
\begin{eqnarray}
\sigma = \frac{3}{4}e,
\end{eqnarray}
independent of $n$. This prediction agrees with the numerically computed growth rates presented in Figs.~\ref{1} and \ref{3} when $e\ll1$. 

This result differs from the result obtained by \cite{John2005a} for a vertically unstructured (i.e.~cylindrical) disc of $\frac{3}{16}e$. We have verified that we also obtain this result by re-deriving Eq.~\ref{growthrate} with $w=g-1=0$, and consider the limit as $n\rightarrow \infty$. The difference between the two predictions arises because of the additional presence of the vertical disc oscillations in an eccentric disc when its vertical structure is considered. This provides an additional free energy source (see Appendix \ref{Energy} below), and an additional periodic forcing that can excite inertial waves. The instability is strongly enhanced and it is essential to consider the vertical structure of the disc to obtain the correct growth rate.

In this case, the resulting phase relation for the pair of waves is $A^{-}_{n}=-\mathrm{i} A^{+}_{n}$, so that the physical instability is a standing wave whose vertical velocity is proportional to (using Eq.~\ref{unstablemode})
\begin{eqnarray}
\mathrm{Re}\left[A^{+}_{n}\mathrm{e}^{-\mathrm{i} \omega \theta+\mathrm{i}k_{\xi}\xi}-A^{-}_{n}\mathrm{e}^{\mathrm{i} \omega \theta+\mathrm{i}k_{\xi}\xi}\right] = 2 |A^{+}_{n}| \sin \left(k_{\xi}\xi+\phi_{A}-\frac{\pi}{4}\right)\sin \left(\omega\theta-\frac{\pi}{4}\right),
\end{eqnarray}
for example, where $\phi_{A}$ is the argument of $A^{+}_{n}$. This consists of the superposition of a pair of travelling waves that propagate radially in opposite directions.

\subsection{Circular reference orbit with a nonzero eccentricity gradient}
For a disc with $e=0$, the growth rate is
\begin{eqnarray}
\sigma = \frac{3}{4}\left(\frac{4n-1}{16n-1}\right)|\lambda e^{\prime}| \;\rightarrow \frac{3}{16} |\lambda e^{\prime}| \;\;\;\; \mathrm{as} \;\;\;\; n\rightarrow \infty.
\end{eqnarray}
The instability of an eccentricity gradient is therefore somewhat weaker than the instability of eccentricity for comparable $e$ and $|\lambda e^{\prime}|$. Neverthless, larger eccentricity gradients might be expected to result from disc-companion tidal interactions. The instability again takes the form of a standing wave, as in the case of a uniformly eccentric disc.

\subsection{Maximum growth rate for a given eccentricity}

The maximum growth rate for a given eccentricity can be estimated by substituting the criterion for the orbits to just intersect (Eq.~\ref{intersect}) into the general expression for the growth rate. Note that the maximum growth rate is obtained for an untwisted disc with the maximum positive eccentricity gradient. In this case we obtain an upper bound on the maximum growth rate, as a function of $e$,
\begin{eqnarray}
\sigma\leq \frac{3}{16}\sqrt{25e^2+10e+1}\rightarrow 1.125, \;\;\;\; \mathrm{as} \;\;\;\; e\rightarrow 1.
\end{eqnarray}
This approximately agrees with the maximum growth rates presented in Fig.~\ref{6}, except for the largest $e$ considered, where this estimate is no longer valid. This indicates that the growth rate for a given eccentricity can be much larger than the corresponding instability in a disc with a uniform eccentricity of the same magnitude.

\section{Energetics of the instability in an isothermal disc}
\label{Energy}
In this section we construct an energy equation from Eqs.~\ref{Eq1a}--\ref{Eq4a}. This will allow us to understand the energetics of the instability analysed in Appendix \ref{Theory}. To construct the energy equation we note that mass conservation requires
\begin{eqnarray}
\partial_{t}(\rho J \Omega)&=&-w J \Omega \partial_{\zeta}(\rho \zeta), 
\end{eqnarray}
and that
\begin{eqnarray}
\partial_{t}g_{ij}&=&\Omega\left(\Gamma^{l}_{ik}g_{lj}+\Gamma^{l}_{jk}g_{il}\right).
\end{eqnarray}
The covariant derivative of a contravariant vector is
\begin{eqnarray}
\label{covariant}
\nabla_{i}v^{j}=\partial_{i}v^{j}+\Gamma^{j}_{ik}v^{k}.
\end{eqnarray}
If we define $U^{i}$ to be the components of the background velocity field (orbital and vertical flow), then
\begin{eqnarray}
\nabla_{\lambda} U^{\lambda}&=&\Gamma^{\lambda}_{\lambda\phi}\Omega, \\
\nabla_{\phi} U^{\lambda}&=&\Gamma^{\lambda}_{\phi\phi}\Omega, \\
\nabla_{\lambda} U^{\phi}&=&\partial_{\lambda}\Omega+\Gamma^{\phi}_{\lambda\phi}\Omega, \\
\nabla_{\phi} U^{\phi}&=&\partial_{\phi}\Omega+\Gamma^{\phi}_{\phi\phi}\Omega, \\
\nabla_{z} U^{z}&=&w.
\end{eqnarray}
We define
\begin{eqnarray}
\mathcal{E}=\frac{1}{2}g_{ij}v^{i}(v^{j})^{*}+\frac{1}{2}|h|^{2},
\end{eqnarray}
to be the specific energy of the perturbations, so that the statement of energy (flux) conservation can be written
\begin{eqnarray}
d_{t}\int_{\infty}^{\infty} \rho J \Omega \mathcal{E} d\zeta &=& -\mathrm{Re}\left\{ \int_{\infty}^{\infty} \rho J \Omega \left[v^{i}(v^{j})^{*}\nabla_{i}U_{j} \right]d\zeta \right\} \\
&=& -\mathrm{Re}\left\{ \int_{\infty}^{\infty} \rho J \Omega \left[v^{i}(v^{j})^{*}g_{jk}\nabla_{i}U^{k} \right]d\zeta \right\}, \\
&=& -\mathrm{Re}\left\{ \int_{\infty}^{\infty} \rho J \Omega \left[ a_{11} |v^{\xi}|^{2} +a_{12}v^{\xi}(v^{\eta})^{*} +a_{22}|v^{\eta}|^{2}+a_{33}|v^{\zeta}|^{2} \right]d\zeta \right\},
\end{eqnarray}
assuming appropriate boundary conditions so that the boundary terms vanish (and noting that the covariant derivative of the metric tensor is zero). We define
\begin{eqnarray}
a_{11}&=& g_{\lambda\lambda}\Gamma^{\lambda}_{\lambda\phi}\Omega+g_{\lambda\phi}(\Gamma^{\phi}_{\lambda\phi}\Omega+\lambda\partial_{\lambda}\Omega) = s\lambda e^{\prime} -c e \lambda \omega^{\prime}-\frac{1}{2}e s +O(\epsilon^2), \\
a_{12}&=& g_{\lambda\lambda}\Gamma^{\lambda}_{\phi\phi}\Omega+g_{\lambda\phi}(\Gamma^{\lambda}_{\lambda\phi}\Omega+\Gamma^{\phi}_{\phi\phi}\Omega+\partial_{\phi}\Omega)+g_{\phi\phi}(\Gamma^{\phi}_{\lambda\phi}\Omega+\lambda\partial_{\lambda}\Omega)=-\frac{3}{2}+(e+2\lambda e^{\prime})c+2 s e\lambda \omega^{\prime} +O(\epsilon^2), \\
a_{22}&=& g_{\lambda\phi}\Gamma^{\lambda}_{\phi\phi}\Omega +g_{\phi\phi}(\Gamma^{\phi}_{\phi\phi}\Omega+\partial_{\phi}\Omega)=-e s +O(\epsilon^2), \\
a_{33}&=& w=3 e s +O(\epsilon^2),
\end{eqnarray}
to be the nonzero components of the background (covariant) velocity gradient tensor. For a uniform circular disc, this reduces to $-\frac{3}{2}v^{\xi}(v^{\eta})^{*}$ on the RHS, as expected. The RHS represents the exchange of energy with the vertical and orbital flows through Reynolds stresses. 

We pose a multiple-scales expansion of the energy equation and consider only terms $O(\epsilon^{1})$ that give a net contribution to the energy of the unstable mode around an orbit, after integrating over $\theta$. After each term is evaluated using the unstable mode written down in Eq.~\ref{unstablemode}, we are left with the following nonzero contributions that result from the left-hand side of the energy equation (before computing the $\zeta$ integral)
\begin{eqnarray}
\label{LHS1}
\partial_{1} \int_{0}^{2\pi} \underbrace{\rho J \Omega \mathcal{E}}_{O(\epsilon^{0})} d\theta = \partial_{1} \int_{0}^{2\pi} \frac{1}{2}\left(|u^{\xi, 0}_{n}|^2 + |u^{\eta,0}_{n}|^2 + \frac{1}{n}|u^{\zeta,0}_{n}|^2+|h_{n}^{0}|^2\right)d\theta=\frac{\pi (4n-1)(20n+1)}{64}\partial_{1}\left(|A^{+}_{n}|^2+|A^{-}_{n}|^2\right),
\end{eqnarray}
where the factor of $(1/n)$ comes from the expansion of the vertical velocity in $\mathrm{He}_{n-1}$ rather than $\mathrm{He}_{n}$, and
\begin{eqnarray}
\label{LHS2}
\int_{0}^{2\pi} \underbrace{\Omega}_{O(\epsilon^{1})} \partial_{0} \underbrace{\rho J \Omega \mathcal{E}}_{O(\epsilon^{0})} d\theta =-\frac{3\pi (4n-1)^2}{32}e\; \mathrm{Im}[A_{n}^{+}(A_{n}^{-})^{*}].
\end{eqnarray}
From the right-hand side of the energy equation, we obtain
\begin{eqnarray}
\label{RHS1}
-\mathrm{Re} \int_{0}^{2\pi} \underbrace{\rho J \Omega}_{O(\epsilon^{0})}\underbrace{a_{11}}_{O(\epsilon^{1})}|u^{\xi,0}_{n}|^2 d\theta=\frac{\pi(4n-1)^2}{64}\left[(-e+2\lambda e^{\prime})\mathrm{Im}[A_{n}^{+}(A_{n}^{-})^{*}]-2e \lambda \omega^{\prime} \mathrm{Re}[A_{n}^{+}(A_{n}^{-})^{*}]\right],
\end{eqnarray}
\begin{eqnarray}
\label{RHS2}
-\mathrm{Re} \int_{0}^{2\pi} \underbrace{\rho J \Omega}_{O(\epsilon^{0})}\underbrace{a_{22}}_{O(\epsilon^{1})}|u^{\eta,0}_{n}|^2 d\theta=\frac{\pi(4n-1)^2}{32}e \;\mathrm{Im}[A_{n}^{+}(A_{n}^{-})^{*}],
\end{eqnarray}
\begin{eqnarray}
\label{RHS3}
-\mathrm{Re} \int_{0}^{2\pi} \underbrace{\rho J \Omega}_{O(\epsilon^{0})}\underbrace{a_{33}}_{O(\epsilon^{1})}(1/n)|u^{\zeta,0}_{n}|^2 d\theta=\frac{9\pi(4n-1)n}{8}e \;\mathrm{Im}[A_{n}^{+}(A_{n}^{-})^{*}],
\end{eqnarray}
\begin{eqnarray}
\label{RHS4}
-\mathrm{Re} \int_{0}^{2\pi} \underbrace{\rho J \Omega}_{O(\epsilon^{1})}\underbrace{a_{12}}_{O(\epsilon^{0})}u^{\xi,0}_{n}(u_{n}^{\eta,0})^{*} d\theta=\frac{9\pi (4n-1)^2 n}{32}e\;\mathrm{Im}[A_{n}^{+}(A_{n}^{-})^{*}],
\end{eqnarray}
\begin{eqnarray}
\label{RHS5}
-\mathrm{Re} \int_{0}^{2\pi} \underbrace{\rho J \Omega}_{O(\epsilon^{0})}\underbrace{a_{12}}_{O(\epsilon^{1})}u_{n}^{\xi,0}(u_{n}^{\eta,0})^{*} d\theta=\frac{\pi (4n-1)^2 }{32}\left[(e+2\lambda e^{\prime})\mathrm{Im}[A_{n}^{+}(A_{n}^{-})^{*}]-2e\lambda \omega^{\prime}\mathrm{Re}[A_{n}^{+}(A_{n}^{-})^{*}]\right],
\end{eqnarray}
\begin{eqnarray}
\label{RHS6}
-\mathrm{Re} \int_{0}^{2\pi} \underbrace{\rho J \Omega a_{12}}_{O(\epsilon^{0})}(u_{n}^{\xi,0}(u_{n}^{\eta,1})^{*}+u_{n}^{\xi,1}(u_{n}^{\eta,0})^{*}) d\theta=-\frac{3\pi (4n-1)^2 }{64}\left[\partial_{1}\left(|A^{+}_{n}|^2+|A^{-}_{n}|^2\right)+(6n+1)e\;\mathrm{Im}[A_{n}^{+}(A_{n}^{-})^{*}]\right].
\end{eqnarray}
For the last term, we must use part of the solution at $O(\epsilon^{1})$. However, it turns out that we only require the relationship between $u_{n}^{\xi,1}$ and $u_{n}^{\eta,1}$, which can be obtained for the equation for $u_{n}^{\eta,1}$, which is:
\begin{eqnarray}
\partial_{0}u_{n}^{\eta,1}+\frac{1}{2}u_{n}^{\xi,1}=-(2ec\partial_{0}+\partial_{1}+n w)u_{n}^{\eta,0}-ecu_{n}^{\xi,0}-2esu_{n}^{\eta,0}+\mathrm{i}k_{\xi} es h_{n}^{0}.
\end{eqnarray}
The components of this equation that we require are those that are proportional to $\mathrm{e}^{\mp \mathrm{i} \theta/2}$:
\begin{eqnarray}
-\mathrm{i}u_{n}^{\eta,1}+u_{n}^{\xi,1}|_{-\mathrm{i}\theta/2}&=&\frac{4n-1}{16}\left[4\partial_{1}A_{n}^{+}+(6n+1)\mathrm{i} e A_{n}^{-}\right], \\
\mathrm{i}u_{n}^{\eta,1}+u_{n}^{\xi,1}|_{\mathrm{i}\theta/2}&=&\frac{4n-1}{16}\left[4\partial_{1}A_{n}^{-}-(6n+1)\mathrm{i} e A_{n}^{+}\right].
\end{eqnarray}
Note also that
\begin{eqnarray}
&&\int_{-\infty}^{\infty}\mathrm{e}^{-\frac{\zeta^2}{2}}\mathrm{He}_{n}(\zeta)\mathrm{He}_{n^{\prime}}(\zeta) d\zeta = n! \sqrt{2\pi}\delta_{nn^{\prime}}, \\
&&\int_{-\infty}^{\infty}\rho J \Omega \left[\mathrm{He}_{n}(\zeta)\right]^{2} d\zeta=\int_{-\infty}^{\infty}J \Omega \mathrm{e}^{f-g \frac{\zeta^2}{2}}\left[\mathrm{He}_{n}(\zeta)\right]^{2} d\zeta = n! \sqrt{2\pi} \left[1-6 n e c\right]+O(\epsilon^2).
\end{eqnarray}

After all this work, we can combine Eqs.~\ref{LHS1}--\ref{RHS6} to obtain an energy equation for the unstable modes:
\begin{eqnarray}
\nonumber
\frac{\pi(4n-1)(16n-1)}{32}\partial_{1}\left(|A^{+}_{n}|^2+|A^{-}_{n}|^2\right)&=&\frac{3\pi(4n-1)}{32}\left[(16n-1) e\;\mathrm{Im}[A_{n}^{+}(A_{n}^{-})^{*}] \right. \\ && \left. \hspace{2cm} +(4n-1)\left(\lambda e^{\prime} \mathrm{Im}[A_{n}^{+}(A_{n}^{-})^{*}] - e\lambda \omega^{\prime} \mathrm{Re}[A_{n}^{+}(A_{n}^{-})^{*}]\right)\right],
\label{EnergyFinal}
\end{eqnarray}
from which we can obtain the growth rate previously written down in Eq.~\ref{growthrate} after looking for growing modes with $A_{n}^{+}, A_{n}^{-}\propto \mathrm{e}^{\sigma \theta_{1}}$. This calculation is useful in two ways. Firstly, it allows us to check Eq.~\ref{growthrate} by providing an alternative derivation of the growth rate of the instability. This is comforting. Secondly, it allows us to determine the primary energy source driving the instability in each case. Note that there is an exact cancellation of terms $O(n^3)$ between Eqs.~\ref{RHS4} and \ref{RHS6}, which would otherwise dominate the right-hand side. The coefficient in square brackets of the first term on the right-hand side of Eq.~\ref{EnergyFinal} is made up of a factor of $4n-1$, which arises even in the absence of laminar flows, and a factor of $12n$ that follows from the inclusion of  the vertical laminar flows (the final two terms in the square brackets, involving the eccentricity gradient, are not affected by the laminar flows for an isothermal disc).

For a uniformly eccentric disc, the largest individual term (at $O(\epsilon^{1})$) that does not cancel is clearly Eq.~\ref{RHS3}, which represents the extraction of energy from the vertical oscillation of the disc. The ratio of this contribution to the total contribution for a uniformly eccentric disc is $12n/(16n-1)\rightarrow 3/4$ for large $n$. The next largest term comes from Eq.~\ref{LHS2}, which represents the amplification of perturbation energy through the time variation of the orbital angular velocity. On the other hand, if the laminar vertical flows in the disc are artificially neglected, Eq.~\ref{RHS3} does not contribute, and the remaining terms give the smaller growth rate obtained by \cite{John2005a} in the limit $n\rightarrow \infty$. This again illustrates the importance of these vertical flows for the instability.

\section{WKB theory of parametric instability in a neutrally stratified polytropic disc}
\label{WKBTheory}
In this section we perform a local stability analysis of an eccentric disc using a WKB approximation, in a neutrally stratified disc with any polytropic (adiabatic) index $n_{p}$. That is, we consider a circular equilibrium disc with $p=K\rho^{1+\frac{1}{n_{p}}}$, where $\gamma=1+\frac{1}{n_{p}}$ for a neutrally stratified (adiabatic) disc. We do not consider stably (or unstably) stratified discs, since the relevant inertial modes are then spatially localised near to the mid-plane \citep{KorycanskyPringle1995,Ogilvie1998}, which requires taking into account their vertical structure, thereby complicating matters. The approach taken here is somewhat similar to that for an isothermal disc presented in Appendix \ref{Theory}, though there are some differences, which will be highlighted below. The calculation in this section is an extension of \cite{John2005a} to take into account an eccentricity gradient, the vertical structure (and oscillations) of the disc, as well as any adiabatic index (for a neutrally stratified disc).

In the WKB approximation, axisymmetric inertial perturbations of an eccentric disc are incompressible (e.g.~\citealt{Ogilvie1998}), and satisfy (cf. Eqs~\ref{Eq1a}--\ref{Eq4a}):
\begin{eqnarray}
\label{Eq1WKB}
&& \Omega\partial_{\theta}v^{\xi}+ w \zeta \partial_{\zeta} v^{\xi}+2\Gamma^{\lambda}_{\lambda \phi} \Omega v^{\xi}+2\Gamma^{\lambda}_{\phi\phi}\Omega v^{\eta}=-g^{\lambda\lambda} \partial_{\xi} h, \\
&& \Omega\partial_{\theta}v^{\eta}+ w \zeta \partial_{\zeta} v^{\eta} +\left(\partial_{\lambda}\Omega+ 2\Gamma^{\phi}_{\lambda \phi} \Omega\right)v^{\xi}+\left(\partial_{\phi}\Omega+2\Gamma^{\phi}_{\phi\phi}\right)v^{\eta} =- \lambda g^{\lambda\phi}\partial_{\xi}h, \\
&&  \Omega\partial_{\theta}v^{\zeta}+ w \zeta \partial_{\zeta} u^{\zeta}+w v^{\zeta}=-\partial_{\zeta}h, \\
&& 0=-\partial_{\xi} v^{\xi}-\partial_{\zeta}v^{\zeta}.
\label{Eq4WKB}
\end{eqnarray}
We analyse the solutions of these equations using Kelvin (shearing) waves with a $\theta$-dependent vertical wavenumber $k_{\zeta}(\theta)$,
\begin{eqnarray}
v^{\xi}=\mathrm{Re}\left[\hat{u}^{\xi}\mathrm{e}^{\mathrm{i}k_{\xi} \xi + \mathrm{i}k_{\zeta}(\theta)\zeta-\mathrm{i} \omega \theta}\right],
\end{eqnarray}
and so on, where we subsequently drop the hats on the perturbations (we also introduce an extra factor of $\lambda^{-1}$ in the $v^{\eta}$ component of the solution so that $\hat{u}^{\eta}$ has units of a velocity). These are locally plane waves with vertical wavelengths that stretch in concert with the vertical oscillations of the disc. Our reason for choosing an evolving vertical wavenumber is to eliminate the terms that are linear in $\zeta$ from Eqs~\ref{Eq1WKB}--\ref{Eq4WKB}, which is accomplished by requiring
\begin{eqnarray}
\Omega \mathrm{d}_{\theta} k_{\zeta}=-w k_{\zeta}.
\end{eqnarray}
The laminar flow solutions no longer satisfy Eq.~\ref{laminar}--\ref{laminar1}, and are instead the solutions of
\begin{eqnarray}
&& (1+e \cos\theta)^2 \mathrm{d}_{\theta}w+w^2=-(1+e\cos\theta)^{3}+g, \\
&& (1+e \cos\theta)^2\mathrm{d}_{\theta}g=-(\gamma-1)\Delta g-(\gamma+1) w g.
\end{eqnarray}
These have the following $2\pi$-periodic solutions:
\begin{eqnarray}
g &=& 1+ \left(\frac{\gamma+1}{\gamma}\right)3ec -\left(\frac{\gamma-1}{\gamma}\right)(c \lambda e^{\prime} +s e \lambda \omega^{\prime})+ O(\epsilon^2), \\
w&=&\frac{3 e s}{\gamma}+\left(\frac{\gamma-1}{\gamma}\right)(-s\lambda e^{\prime} +c e \lambda \omega^{\prime}) + O(\epsilon^2), \\
k_{\zeta}&=& k_{\zeta}^{0}\left[1+\frac{3ec}{\gamma} - \left(\frac{\gamma-1}{\gamma}\right)(c \lambda e^{\prime} +s e \lambda \omega^{\prime})\right] + O(\epsilon^2),
\end{eqnarray}
which reduce to the solutions obtained in Appendix \ref{Theory} for an isothermal disc when $\gamma=1$. Note, that an eccentricity gradient plays a role in driving these oscillations when $\gamma\ne1$, unlike for the case of an isothermal disc.

We define a small parameter $\epsilon$ such that $e$, $|\lambda e^{\prime}|$  and $e|\lambda \omega^{\prime}|$ are each $O(\epsilon)$, and use a multiple-time-scales expansion as in Appendix \ref{Theory}. We consider an instability of a pair of inertial waves with $\omega=\pm\frac{1}{2}$, with a single vertical wavenumber $k^{0}_{\zeta}=n$ (where $n$ is used as a label for the mode), which can be written as
\begin{eqnarray}
\boldsymbol{U}_{n}^{0}=A^{+}_{n}(\theta_{1})\hat{\boldsymbol{U}}^{+0}_{n}+A^{-}_{n}(\theta_{1})\hat{\boldsymbol{U}}^{-0}_{n},
\end{eqnarray}
where $\boldsymbol{U}_{n}=\left[u^{\xi}_{n},u^{\eta}_{n},u^{\zeta}_{n},h_{n}\right]^{T}$, and the eigenvectors are 
\begin{eqnarray}
\hat{\boldsymbol{U}}^{\pm0}_{n} = \left[ \begin{array}{c}
\pm \mathrm{i}\omega \\
\frac{1}{2} \\
\mp \mathrm{i}\frac{k_{\xi}}{k_{\zeta}}\omega \\
-\mathrm{i} \frac{k_{\xi}}{k_{\zeta}^{2}} \omega^{2}
 \end{array} \right] \mathrm{e}^{\mp \mathrm{i}\omega \theta_{0}}.
\end{eqnarray}
The corresponding system at $O(\epsilon^{0})$
\begin{eqnarray}
L_{n}\boldsymbol{U}^{0}_{n}=\boldsymbol{0},
\end{eqnarray}
where
\begin{eqnarray}
L_{n} = \left( \begin{array}{cccc}
\partial_{0} & -2 & 0 & \mathrm{i}k_{\xi} \\
\frac{1}{2} & \partial_{0} & 0  & 0\\
0 & 0 & \partial_{0} & \mathrm{i} k^{0}_{\zeta} \\
k_{\xi} & 0 & k^{0}_{\zeta} & 0 \end{array}
\right),
\end{eqnarray}
The associated solvability condition for a general forcing vector $\boldsymbol{F}$ is:
\begin{eqnarray}
-\mathrm{i}\omega a_{n}+2 b_{n}+\mathrm{i}\omega \frac{k_{\xi}}{k_{\zeta}}c_{n}-\frac{k_{\xi}}{k_{\zeta}^{2}}\omega^{2} d_{n}=0,
\end{eqnarray}
which allows us to obtain the amplitude equations for the two waves at $O(\epsilon^{1})$:
\begin{eqnarray}
\partial_{1}A_{n}^{\pm}=\mp \frac{3}{16\gamma}\left[(3+\gamma)e+\lambda e^{\prime} \pm \mathrm{i}e\lambda\omega^{\prime}\right]A_{n}^{\mp}.
\end{eqnarray}
The growth rate of the instability at exact resonance is therefore
\begin{eqnarray}
\sigma &=& \frac{3}{16\gamma} \sqrt{(3+\gamma)^2 e^2 + (\lambda e^{\prime})^2 +2(3+\gamma) e \lambda e^{\prime} +e^2(\lambda\omega^{\prime})^{2}}, \\
&=& \frac{3}{16\gamma} |\left(3+\gamma\right) E+\lambda E^{\prime}|.
\end{eqnarray}
This is equivalent to the result for the isothermal disc (Eq.~\ref{growthrate}, when $n\rightarrow \infty$).
Note that when $\gamma=\frac{5}{3}$, the instability of a uniformly eccentric disc has a growth rate $\sigma=\frac{21}{40}e\approx 0.525 e$, which is somewhat smaller than the isothermal (most compressible case) but is still significantly enhanced over the case in which the vertical structure of the disc is neglected ($\frac{3}{16}e$).

We can conclude from this that taking into account the vertical structure of the disc can significantly amplify the growth rate for any realistic adiabatic index with $\gamma\approx 1.4-1.7$. The strongest amplification is clearly for the isothermal disc; however, the vertical oscillations of the disc play an important role for any realistic adiabatic index.

\subsection{Incompressible limit}
The incompressible ($\gamma\rightarrow\infty$) limit of the equations gives the laminar solutions
\begin{eqnarray}
&& w=-\Delta, \\
&& g=(1+e \cos\theta)^{3}+\Delta^{2}-(1+e \cos\theta)^{2}\mathrm{d}_{\theta}\Delta,
\end{eqnarray}
where $\Delta$ is known in advance based on the local properties of the Keplerian orbit. The thickness of disc varies only if there is a nonzero orbital velocity divergence -- whenever the orbital streamlines bunch up, this forces the disc to become thicker since the fluid is incompressible. The instability in this case has growth rate
\begin{eqnarray}
\sigma &=& \frac{3}{16} e,
\end{eqnarray}
which is independent of the local eccentricity gradient.

\setlength{\bibsep}{0pt}
\bibliography{disc}

\begin{thebibliography}{}

\bibitem[\protect\citeauthoryear{{Barker} \& {Lithwick}}{{Barker} \&
  {Lithwick}}{2013}]{BL2013a}
{Barker} A.~J.,  {Lithwick} Y.,  2013, MNRAS, 435, 3614

\bibitem[\protect\citeauthoryear{{Barker} \& {Lithwick}}{{Barker} \&
  {Lithwick}}{2014}]{BL2013b}
{Barker} A.~J.,  {Lithwick} Y.,  2014, MNRAS, 437, 305

\bibitem[\protect\citeauthoryear{{Bitsch}, {Crida}, {Libert} \&
  {Lega}}{{Bitsch} et~al.}{2013}]{Bitsch2013}
{Bitsch} B.,  {Crida} A.,  {Libert} A.-S.,    {Lega} E.,  2013, A\&A, 555, A124

\bibitem[\protect\citeauthoryear{{D'Angelo}, {Lubow} \& {Bate}}{{D'Angelo}
  et~al.}{2006}]{DAngelo2006}
{D'Angelo} G.,  {Lubow} S.~H.,    {Bate} M.~R.,  2006, ApJ, 652, 1698

\bibitem[\protect\citeauthoryear{{Goldreich} \& {Sari}}{{Goldreich} \&
  {Sari}}{2003}]{GoldreichSari2003}
{Goldreich} P.,  {Sari} R.,  2003, ApJ, 585, 1024

\bibitem[\protect\citeauthoryear{{Goodman}}{{Goodman}}{1993}]{Goodman1993}
{Goodman} J.,  1993, ApJ, 406, 596

\bibitem[\protect\citeauthoryear{{Kerswell}}{{Kerswell}}{2002}]{Kerswell2002}
{Kerswell} R.~R.,  2002, Annual Review of Fluid Mechanics, 34, 83

\bibitem[\protect\citeauthoryear{{Kley} \& {Dirksen}}{{Kley} \&
  {Dirksen}}{2006}]{KleyDirksen2006}
{Kley} W.,  {Dirksen} G.,  2006, A\&A, 447, 369

\bibitem[\protect\citeauthoryear{{Kley}, {Papaloizou} \& {Ogilvie}}{{Kley}
  et~al.}{2008}]{KleyPO2008}
{Kley} W.,  {Papaloizou} J.~C.~B.,    {Ogilvie} G.~I.,  2008, A\&A, 487, 671

\bibitem[\protect\citeauthoryear{{Korycansky} \& {Pringle}}{{Korycansky} \&
  {Pringle}}{1995}]{KorycanskyPringle1995}
{Korycansky} D.~G.,  {Pringle} J.~E.,  1995, MNRAS, 272, 618

\bibitem[\protect\citeauthoryear{{Lubow}}{{Lubow}}{1991a}]{Lubow1991a}
{Lubow} S.~H.,  1991a, ApJ, 381, 259

\bibitem[\protect\citeauthoryear{{Lubow}}{{Lubow}}{1991b}]{Lubow1991b}
{Lubow} S.~H.,  1991b, ApJ, 381, 268

\bibitem[\protect\citeauthoryear{{Lubow}, {Pringle} \& {Kerswell}}{{Lubow}
  et~al.}{1993}]{LubowPringleKerswell1993}
{Lubow} S.~H.,  {Pringle} J.~E.,    {Kerswell} R.~R.,  1993, ApJ, 419, 758

\bibitem[\protect\citeauthoryear{{Marzari}, {Baruteau}, {Scholl} \&
  {Thebault}}{{Marzari} et~al.}{2012}]{Marzari2012}
{Marzari} F.,  {Baruteau} C.,  {Scholl} H.,    {Thebault} P.,  2012, A\&A, 539,
  A98

\bibitem[\protect\citeauthoryear{{Ogilvie}}{{Ogilvie}}{1998}]{Ogilvie1998}
{Ogilvie} G.~I.,  1998, MNRAS, 297, 291

\bibitem[\protect\citeauthoryear{{Ogilvie}}{{Ogilvie}}{2001}]{Ogilvie2001}
{Ogilvie} G.~I.,  2001, MNRAS, 325, 231

\bibitem[\protect\citeauthoryear{{Ogilvie}}{{Ogilvie}}{2008}]{Ogilvie2008}
{Ogilvie} G.~I.,  2008, MNRAS, 388, 1372

\bibitem[\protect\citeauthoryear{{Ogilvie} \& {Barker}}{{Ogilvie} \&
  {Barker}}{2014}]{OB2014a}
{Ogilvie} G.~I.,  {Barker} A.~J.,  2014, MNRAS

\bibitem[\protect\citeauthoryear{{Ogilvie} \& {Latter}}{{Ogilvie} \&
  {Latter}}{2013}]{OL2013b}
{Ogilvie} G.~I.,  {Latter} H.~N.,  2013, MNRAS, 433, 2420

\bibitem[\protect\citeauthoryear{{Okazaki}}{{Okazaki}}{1991}]{Okazaki1991}
{Okazaki} A.~T.,  1991, PASJ, 43, 75

\bibitem[\protect\citeauthoryear{{Okazaki}, {Kato} \& {Fukue}}{{Okazaki}
  et~al.}{1987}]{Okazaki1987}
{Okazaki} A.~T.,  {Kato} S.,    {Fukue} J.,  1987, PASJ, 39, 457

\bibitem[\protect\citeauthoryear{{Papaloizou}}{{Papaloizou}}{2002}]{Pap2002}
{Papaloizou} J.~C.~B.,  2002, A\&A, 388, 615

\bibitem[\protect\citeauthoryear{{Papaloizou}}{{Papaloizou}}{2005a}]{John2005a}
{Papaloizou} J.~C.~B.,  2005a, A\&A, 432, 743

\bibitem[\protect\citeauthoryear{{Papaloizou}}{{Papaloizou}}{2005b}]{John2005b}
{Papaloizou} J.~C.~B.,  2005b, A\&A, 432, 757

\bibitem[\protect\citeauthoryear{{Papaloizou}, {Nelson} \&
  {Masset}}{{Papaloizou} et~al.}{2001}]{PapNM2001}
{Papaloizou} J.~C.~B.,  {Nelson} R.~P.,    {Masset} F.,  2001, A\&A, 366, 263

\bibitem[\protect\citeauthoryear{{Papaloizou}, {Savonije} \&
  {Henrichs}}{{Papaloizou} et~al.}{1992}]{PapSav1992}
{Papaloizou} J.~C.~B.,  {Savonije} G.~J.,    {Henrichs} H.~F.,  1992, A\&A,
  265, L45

\bibitem[\protect\citeauthoryear{{Peiris} \& {Tremaine}}{{Peiris} \&
  {Tremaine}}{2003}]{PeirisTremaine2003}
{Peiris} H.~V.,  {Tremaine} S.,  2003, ApJ, 599, 237

\bibitem[\protect\citeauthoryear{{Ryu} \& {Goodman}}{{Ryu} \&
  {Goodman}}{1994}]{RyuGoodman1994}
{Ryu} D.,  {Goodman} J.,  1994, ApJ, 422, 269

\bibitem[\protect\citeauthoryear{{Smith}, {Haswell}, {Murray}, {Truss} \&
  {Foulkes}}{{Smith} et~al.}{2007}]{Smith2007}
{Smith} A.~J.,  {Haswell} C.~A.,  {Murray} J.~R.,  {Truss} M.~R.,    {Foulkes}
  S.~B.,  2007, MNRAS, 378, 785

\bibitem[\protect\citeauthoryear{{Tremaine}}{{Tremaine}}{1995}]{Tremaine1995}
{Tremaine} S.,  1995, AJ, 110, 628

\bibitem[\protect\citeauthoryear{{Tremaine}}{{Tremaine}}{2001}]{Tremaine2001}
{Tremaine} S.,  2001, AJ, 121, 1776

\bibitem[\protect\citeauthoryear{{Whitehurst}}{{Whitehurst}}{1988}]{Whitehurst1988}
{Whitehurst} R.,  1988, MNRAS, 232, 35

\end{thebibliography}
\bibliographystyle{mn2e}
\label{lastpage}
\end{document}